%% file: 0_AKG.tex
\documentclass[sigconf,authorversion,nonacm]{acmart}

\usepackage{hologo}
\usepackage{multirow}
\usepackage{subfigure}
\usepackage{xspace}
\usepackage{pifont}
\usepackage{makecell}
\usepackage{xcolor}
\usepackage{booktabs}
\usepackage{siunitx}
\usepackage{tabularx}
\usepackage{placeins}
\usepackage[htt]{hyphenat}
\usepackage[ruled,linesnumbered]{algorithm2e}

\newif\ifSimpleMode\SimpleModefalse

\newcommand{\cmark}{\ding{51}}%
\newcommand{\xmark}{\ding{55}}%

\newcommand{\SysName}{MultiKG\xspace}

\renewcommand{\texttt}[1]{%
  \begingroup
  \ttfamily
  \begingroup\lccode`~=`/\lowercase{\endgroup\def~}{/\discretionary{}{}{}}%
  \begingroup\lccode`~=`[\lowercase{\endgroup\def~}{[\discretionary{}{}{}}%
  \begingroup\lccode`~=`.\lowercase{\endgroup\def~}{.\discretionary{}{}{}}%
  \catcode`/=\active\catcode`[=\active\catcode`.=\active
  \scantokens{#1\noexpand}%
  \endgroup
}

\AtBeginDocument{%
  \providecommand\BibTeX{{%
    \normalfont B\kern-0.5em{\scshape i\kern-0.25em b}\kern-0.8em\TeX}}}

\begin{document}

\SimpleModefalse



\title{\SysName: Multi-Source Threat Intelligence Aggregation for High-Quality Knowledge Graph Representation of Attack Techniques}




\begin{abstract}
  \input{1_Abstract}
\end{abstract}




\author{Jian Wang}
\affiliation{%
  \institution{College of Computer Science and Technology, Zhejiang University}
  \city{Hangzhou}
  \postcode{310027}
  \country{China}
}
\email{wangjian1998@zju.edu.cn}

\author{Tiantian Zhu}
\affiliation{%
  \institution{College of Computer Science and Technology, Zhejiang University of Technology}
  \city{Hangzhou}
  \postcode{310027}
  \country{China}
}
\email{ttzhu@zjut.edu.cn}

\author{Chunlin Xiong}
\affiliation{%
  \institution{China Unicom (Guangdong) Industrial Internet Co., Ltd.}
  \city{Guangdong}
  \postcode{510000}
  \country{China}
}
\email{chunlinxiong@gmail.com}

\author{Yan Chen}
\affiliation{%
  \institution{Department of Computer Science, Northwestern University}
  \city{Evanston}
  \state{Illinois}
  \postcode{60208}
  \country{USA}
}
\email{ychen@northwestern.edu}


\maketitle
\thispagestyle{plain}  
\pagestyle{plain}  
\input{2_Introduction}

\input{3_Background}

\input{4_System_Overview}
\input{5_Threat_Knowledge_Gathering}
\input{7_Evaluation}
\input{9_Related_Wrok}
\input{10_Conclusion}

\bibliographystyle{ACM-Reference-Format}
\bibliography{AKG}
\appendix
\input{11_Appendix}

\end{document}

%% file: 1_Abstract.tex
The construction of attack technique knowledge graphs aims to transform different types of attack knowledge into structured representations that can be used to represent the attack procedure more effectively. Although previous studies have proposed various methods for constructing attack knowledge graphs, these methods generally limit their knowledge sources to textualized data such as Cyber threat intelligence (CTI) reports, which are coarse-grained and unstructured, making the knowledge graphs constructed based on them incomplete and inaccurate.

To address these limitations, we attempt to expand the attack knowledge sources by introducing audit logs and static code analysis on top of CTI reports to provide finer-grained knowledge for constructing attack technique knowledge graphs.
Therefore, we propose a fully automated framework across threat knowledge sources called \SysName automates the processing of data from different sources generates knowledge graphs separately, and then builds a unified attack knowledge graph representation.
Through system design and the utilization of the Large Language Model (LLM), \SysName automates the analysis, construction, and merging of attack graphs from CTI reports, dynamic logs, and static code, respectively. \SysName then aggregates the attack graphs across sources and merges them into a fine-grained unified attack technology knowledge graph representation that encompasses multiple sources.

We implemented and deployed \SysName, then evaluated it using 1015 real attack techniques, and corresponding 9006 attack technique intelligence from real-world CTI reports.
The results show that \SysName is capable of accurately extracting attack knowledge graphs from different sources, and efficiently aggregating and summarizing technical attack knowledge across different resources to generate accurate and complete technique attack knowledge graphs. We also show through concrete case studies that our attack knowledge graph directly benefits downstream security practices tasks such as attack reconstruction and attack detection.

%% file: 2_Introduction.tex
\section{Introduction}




    

Cyber attacks have evolved rapidly in recent years, using increasingly advanced, sophisticated, and versatile attack tactics and techniques to construct covert attack operations, making intrusion detection increasingly challenging. 
A publicly-available test conducted by MITRE ATT\&CK \cite{mitre} Evaluation on intrusion detection products from top security companies such as Symantec \cite{Symantec} and Crowd Strike \cite{CrowdStrike} shows that most products would be bypassed by Live-off-the-Land, attack variants, and other advanced attack methods to bypass. 

To combat the rapidly evolving threat landscape, security practitioners are actively sharing and collecting intelligence of real attack cases, including structured Indicators of Compromises (IOCs), unstructured attack analysis reports, and so on, on public platforms. 
However, collecting and summarizing intelligence from the heterogeneous multi-source platforms is time- and labor-expensive. 
Moreover, the question of how to integrate data from different sources, formats, and representations so that it can represent attack knowledge more accurately and efficiently has not yet been addressed.
(a) \textit{How to collect and summarize this multi-source and heterogeneous threat knowledge to accurately represent and cover complex attack variants?} and (b) \textit{How to effectively integrate this threat knowledge to represent attacks in a more efficient and unified way?} are still open questions that require further research. 



Table~\ref{tab:related_work} list existing.
We list relevant threat intelligence analysis efforts in Table \ref{tab:related_work}. For Extractor~\cite{Satvat2021}, ThreatRap~\cite{gao}, and AttacKG \cite{li2022attackg}, they try to summarize information in threat intelligence to more effectively summarize attack-related knowledge and use it to reconstruct attacks. However, because it does not consider the correlation between security threat intelligence content and real attack logs, it cannot be used for attack detection, but only for a more complete summary of attack-related knowledge. Then, for TTPDrill~\cite{Husari2017} and rcATT~\cite{legoy2020automated}, they only focus on single entity information without considering the correlation between these nodes, and the lack of necessary structural and semantic information, which makes it easy to generate false positives and a low detection rate. In addition, for the work of Poirot~\cite{Milajerdi2019}, iACE~\cite{Liao}, etc., they use the complete attack process in the report as a graph to detect the attack, which makes it easy to bypass by the extension of the attack chain and the replacement of the attack technology.

More importantly, all the articles mentioned in the table, use a single threat intelligence information as the information source, that is, structured and unstructured text, and do not consider other information sources except text descriptions, which makes their representation of the attack very coarse-grained, and the effect of practical application is limited.
This reliance on a single source of information fails to capture a broader attack surface and complex threat patterns. In real-world environments, attackers often utilize multiple means and vectors to execute attacks, and these behaviors are dispersed and recorded in different sources, such as threat reports, audit logs, and attack codes, which are equally critical and can provide more granular threat intelligence. Therefore, future research should explore ways to integrate multiple sources of information to build a more detailed and comprehensive threat intelligence mapping in order to improve threat detection and response capabilities.

In summary, the threat intelligence graph extracted from existing work has the following limitations:
\textbf{1)} Attack knowledge is dispersed across multiple sources, and individual sources usually focus on limited/incomplete attack representations, which do not allow for a complete restoration of the attack.
\textbf{2)} There is a lack of fine-grained information about the underlying attack behavior, such as audit logs during the attack process and information about the attack execution code, which leads to a lack of fine-grained semantic information about the attack semantics in its representation of the attack technique, and prevents it from accurately representing the attack.

The \SysName we propose combines three sources of information: threat intelligence, static code analysis, and dynamic log analysis. \SysName can realize the automatic structured representation of information from different sources, as well as the fusion and summary of attack knowledge from different sources. 
We refine the granularity to the graph structure level of single atomic attack techniques, which can accurately restore the attack graph structure, thus realizing real attack reconstruction at the level of technique.

\begin{table}[htbp]
\scriptsize
    \centering
    \caption{Comparison with cyber threat intelligence gathering approaches}
    \vspace{-0.11in}
    \begin{tabular}{cccccc}
        \toprule
         & \makecell[c]{Mapping-to\\-audit-logs} & \makecell[c]{Graph\\-structure} & \makecell[c]{Technique\\-aware} & \makecell[c]{Cross\\-sources} \\
         \midrule
        Poirot~\cite{Milajerdi2019} & \cmark & \cmark & \xmark & \xmark \\
        iACE~\cite{Liao} & \cmark & \xmark & \xmark & \xmark \\
        Extractor~\cite{Satvat2021} \& ThreatRaptor~\cite{gao}  & \xmark & \cmark & \xmark & \xmark\\
        TTPDrill~\cite{Husari2017} \& rcATT\cite{legoy2020automated}, etc. & \cmark & \xmark & \cmark & \xmark\\
        AttacKG \cite{li2022attackg} & \xmark & \cmark & \cmark & \xmark\\
        \SysName & \cmark & \cmark & \cmark & \cmark\\
        \bottomrule
    \end{tabular}
    \label{tab:related_work}
\end{table}

The main challenge comes from:
\begin{list}
{\labelitemi}{\leftmargin=1.5em}
    \item How to collect and summarize the attack information from different sources(including CTI reports, static code, and dynamically executed audit logs) and handle the attack data separately to get the representation of the attack technique in different sources.
    \item How to integrate the information from different sources and different granularities, fusing multiple sources to get accurate and complete coverage of the attack techniques.
\end{list}
To overcome these challenges, we proposed \SysName, the first cross-source threat intelligence collection and aggregation system. 
In summary, this paper makes the following contributions:

\begin{list}
{\labelitemi}{\leftmargin=1.5em}

    \item Multi-source threat knowledge gathering and aggregation. We proposed the first framework \SysName in this paper for aggregation of knowledge from different threat intelligence sources with respective advantages. Specifically, we combine three sources of information: threat intelligence, static code analysis, and dynamic log analysis. Dynamic logs can provide the topology (including edge and node information) of the actual execution process of attack techniques. Static codes can help complete missing nodes and edges in dynamic graphs. Threat intelligence can help us obtain different variants of a certain attack technology as much as possible so that our final representation of a single technology is not only accurate but also universal.
    \item We design the complete \SysName system to realize the above details. Specifically, we propose an effective algorithm to extract attack technique graphs from huge audit logs, we obtain attack nodes from static code by abstracting syntax trees, and we utilize LLM to analyze entities, entity types, and relationships in threat reports to obtain attack knowledge graphs. We also design single-resource fusion and multi-resource merging algorithms to process the above data to achieve a complete and unified knowledge representation at the attack technique level.
    \item We implemented \SysName, then evaluated it with 1015 real attack techniques and corresponding 9006 attack technique intelligence from real-world CTI reports. The results show that \SysName can accurately extract attack knowledge graphs from different sources and efficiently summarize and aggregate technical-level attack knowledge across resources to generate technical-level accurate and complete attack representations. We also show the benefits of \SysName through specific case studies.
\end{list}

%% file: 3_Background.tex
\section{Motivation \& Limitation}
In this subsection, we present the shortcomings of existing tools, as well as a motivating example to illustrate the advantages of the threat knowledge presented in this paper over existing work.










    

\subsection{Existing Tools Limitations}
At present, the existing attack investigation system has two general ideas: one is to detect attacks by learning normal events in the system through anomaly detection. The other is to match intrusion attacks by learning attack knowledge in threat reports.

Anomaly detection methods can detect unknown attacks but require a stable system environment, otherwise, a large number of false positives may be generated and need to be relearned. And it may be bypassed by super nodes or imitation attacks.

Using real attack data provided by threat intelligence can effectively construct attack templates and conduct attack investigations. However, due to the limited information that threat reports written in natural language can provide, only coarse-grained attack knowledge graphs can be generated, which will affect the efficiency and accuracy of matching, and there is an urgent need to improve it by adding newer information sources.

\subsection{Motivating Example}
\begin{figure*}[h!]
    \centering
    \includegraphics[width=0.75\textwidth]{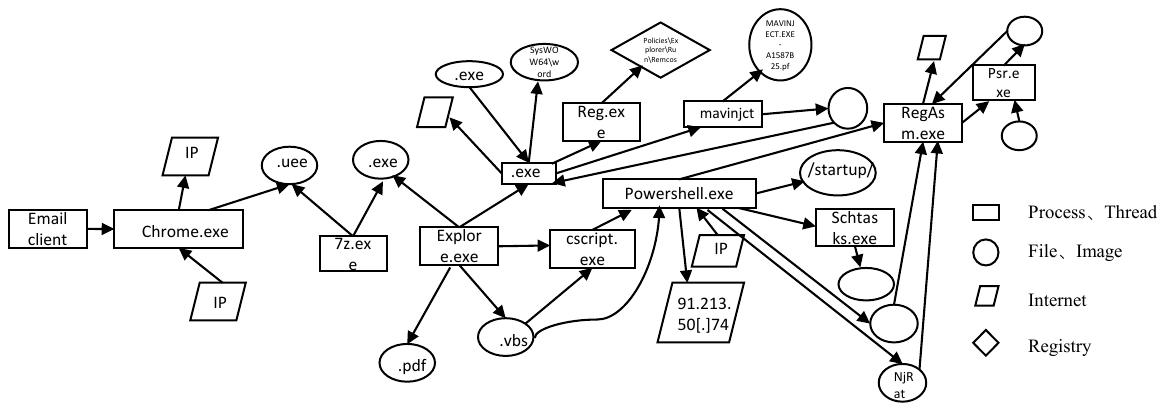}
        \caption{Complete Attack Flow of APT-C-36 Reproduced Based on the Report.}
    \label{fig:Motivation}
\end{figure*}

As shown in Figure~\ref{fig:Motivation}, we reconstructed the complete attack graph based on the spear-phishing attack described in the latest APT-C-36 attack report~\cite{APT-C-36-cnsec, APT-C-36-BlackBerry2023, APT-C-36-360Threat}.
We have supplemented and refined the graph by incorporating cross-source knowledge. The entire attack process involves seven distinct attack techniques.

\begin{figure*}[h!]
    \centering    \includegraphics[width=0.65\textwidth]{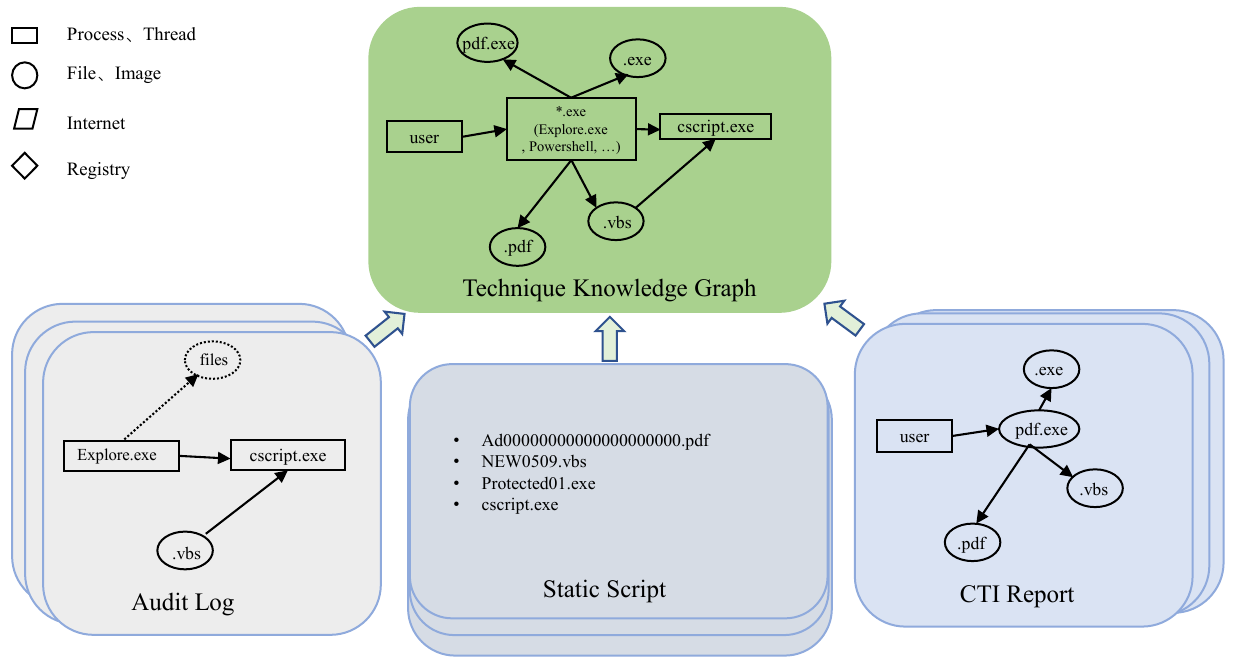}
        \caption{Multi-Source Information Aggregation on Attack Technique \texttt{T1059.005}}
    \label{fig:MotivationSingle}
\end{figure*}

However, if threat intelligence is obtained from a single source alone, for each attack technique, we can only get extremely limited and incomplete entity information from that report, and its representation of inter-entity relationships will ignore some of the nodes and edges in the detailed execution process, leading to the incompleteness of its construction. Secondly, due to its single source of data, it can also not cover different variant types about the same attack technique, further making it incomplete and inaccurate.
This makes attack graphs that only obtain their knowledge sources from reports ineffective for practical application.

As shown in Figure~\ref{fig:MotivationSingle}, we take one of the attack techniques \texttt{T1059.005}, and use it as a specific example to illustrate the benefits of cross-source knowledge fusion.
Through the report, we can only learn that the attacker utilizes \texttt{pdf.exe} to release \texttt{.vbs} to launch the attack, but we lack the specific execution details of it and can only use it to build a coarse-grained attack graph.
Through the audit log, we were able to get the real execution process of the event, the attacker used \texttt{Explore.exe} to execute the \texttt{pdf} file and then started the \texttt{cscript.exe} process to load and execute the \texttt{.vbs} attack script, thereby executing the next attack technique. (Using \texttt{powershell.exe} to modify the system startup folder and \texttt{Reg.exe} to modify the registry, thereby achieving persistent operations.) 
By analyzing the static code, we extracted a complete list of entities involved in the attack, such as \texttt{NEW0509.vbs}, \texttt{ad00000000000000000000000.pdf}, etc., and supplemented and refined the relevant nodes and edges in the log-based dynamic graph. In other words, static code analysis provided valuable information for selecting appropriate events and nodes within the extensive audit logs.
Finally, through the fusion of multi-source information, we obtained a complete, fine-grained knowledge graph representation of the \texttt{T1059.005} attack technique.

We combine static and dynamic analysis to effectively fill the information gap in the CIT report. Note that in our actual implementation, there is more than one audit log and threat intelligence description of a particular attack technique, and we perform the merging algorithm on its content before cross-source merging. We will introduce this part in detail in \S \ref{S4} and \S \ref{S5}.

The collection and aggregation of multi-source threat intelligence makes the attack technique graph we finally obtain more complete and fine-grained, more similar to real attacks, and even covers different variants of the same technique.
This enables it to be actually used in the detection and reconstruction work of real attacks, such as reproducing the work of Figure~\ref{fig:MotivationSingle} or detecting and recognizing different attack techniques in it.
We will explain this in more detail in \S\ref{Evaluation} with the case study.

%% file: 4_System_Overview.tex
\section{Preliminary}
\subsection{Threat Model}
We assume the underlying OS, the auditing engine, and monitoring data to be part of the trusted computing base (TCB). Attacks that can compromise these two auditing systems are beyond the scope of this study. This assumption is consistent with previous literature \cite{milajerdi2019holmes}\cite{xiong2020conan}\cite{yang2020uiscope}\cite{zeng2021watson}.

\subsection{Definition of Attack Knowledge Graph}

We want to find an event set E, where each event e is related to an attack, thus forming a complete attack subgraph.

An event $E$ can be defined as a triple $(Subject, Relation, Object)$, where the Subject is a Process entity, the Object is a system entity (ie registry, file, or network socket), and the Relation is the action that the subject points to the object.

A provenance graph is defined as $Graph=(V, E)$, where the node set $V$ represents all system entities and the edge set $E$ represents all system events. Each edge $E=(Subject, Relation, Object)$ denotes that system entity $Subject$ performs operation $Relation$ on system entity $Object$.

The attack technique graph is a subgraph of the provenance graph and represents the complete set of nodes and edges involved in executing a specific attack technique. 
For example, Figure~\ref{fig:Motivation} is the provenance graph of a complete attack process involving all the attack steps. (which we call Tactics). Figure~ \ref{fig:MotivationSingle} shows the technique responsible for the execution of the \texttt{Command and Scripting Interpreter}.

For an attack knowledge graph, we first need to filter the attack-related entities and relationships from different sources of data that constitute the attack events and do the analysis of the attack-related events.
After all the components $(subject, object, and relationship)$ are identified and categorized, we start constructing the attack knowledge graph. This is a structured representation where attack technique-related nodes represent entities $(subject or object)$ and edges represent relationships between them and the whole graph represents the complete execution of an attack technique.

To enhance the richness of the knowledge graph, we integrate additional contextual information such as threat intelligence reports, audit logs, and static code. This enrichment helps to place events in the broader context of cyber threats, making the knowledge graph more accurate and all-encompassing.

By following the above principles, we have developed a framework for building comprehensive and accurate attack knowledge graphs.
It provides security analysts with a clearer and more detailed understanding of attacker behavior, enabling them to design robust defense strategies and enhance the overall security posture. In addition, this graph facilitates more effective detection and response to attacks, as well as threat analysis.

\subsection{Data Collection}
To gather large-scale audit log data encompassing APT techniques from ATT\&CK~\cite{mitre}, we leverage Atomic Red Team~\cite{AtomicRedTeam}, a security testing tool by Red Canary that provides code implementations for various attack techniques, as well as an automated execution framework $\footnote{https://github.com/redcanaryco/atomic-red-team}$ based on Atomic Red Team. We also use our ETW-based~\cite{ETW} automated data collector to capture audit logs.

Specifically, each script provided by the Atomic Red Team covers a specific APT technique in ATT\&CK. To simulate the APT techniques, we created a secure virtual environment on Windows using VMware and then used PowerShell tools to satisfy the attack pre-conditions and fully execute the scripts by granting them system privileges. 
Then, the provenance graph data was collected through an automated ETW-based data collector that we designed.
Each collected event is a triple $(subject, operation, object)$ and is modeled as two nodes connected by an edge of a specific relationship.

Additionally, the Atomic Red Team provided the execution source code for each attack technique, which we parsed using an abstract syntax tree to extract the required node information.

Finally, ATT\&CK provides multiple descriptions from different reports about each attack technique, and we benefit from Openai's GPT-4 API to analyze this threat intelligence and get the attack threat graph.

We utilized the source code and executable frameworks of the 1015 attack procedures of the 282 attack techniques crawled from the atomic red team as the source of the base dataset.
Based on the technique numbers corresponding to these attack techniques in the atomic red team, we crawled all the threat report procedure descriptions corresponding to a single attack technique from the technique details page in MITRE ATT\&CK, respectively, totaling 9006 attack report descriptions.

\subsection{System Architecture}
\begin{figure*}[h!]
    \centering
    \includegraphics[width=0.9\textwidth]{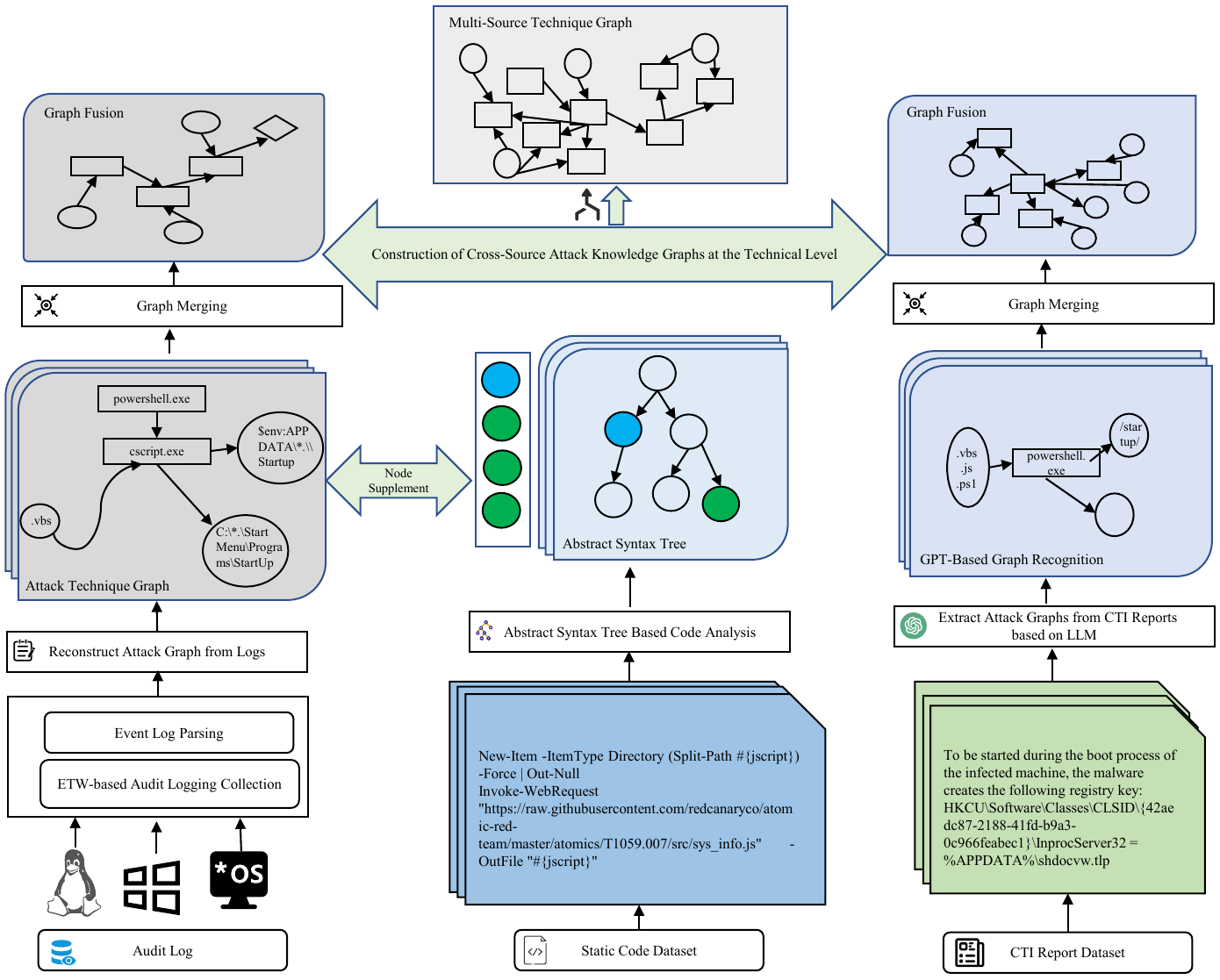}
    \caption{Overview of \SysName Architecture}
    \label{fig:Overview}
\end{figure*}

Figure~\ref{fig:Overview} shows the architecture of 
\SysName. Overall, \SysName mainly consists of two subsystems: 1) single-source threat knowledge analysis module: used for data parsing and attack graph construction for each threat knowledge data source separately; 2) multi-source attack knowledge graph analysis module: used for internal aggregation of single-source attack graphs and aggregation of different cross-source attack graphs in order to realize a unified representation of cross-source attack knowledge graphs.

In \S\ref{S4}, we introduce how to deal with (1) System Audit Logs (\S\ref{SLOG}), (2) Static Code (\S\ref{SAST}), and (3) Cyber Threat Intelligence Report (\S\ref{SCTI}), respectively, and in \S\ref{S5}, we introduce how to do internal aggregation for single-source attack graphs (\S\ref{S51merghsingle}) and cross-source aggregation (\S\ref{S52}). We experimentally validate the above two modules in \S\ref{Evaluation} to show the effectiveness of our system.


%% file: 5_Threat_Knowledge_Gathering.tex
\section{Threat Knowledge Extraction from Different Source}\label{S4}
In this section, we systematically describe how attack data is collected and analyzed from different sources to achieve an independent representation of attack knowledge within each source.

In \S\ref{SLOG}, we describe how to restore attack graphs from audit logs, in \S\ref{SAST}, we show the process of obtaining information about attack nodes via ASTs, and in \S\ref{SCTI}, we show how to analyze the information reported by the MITRE ATT\&CK and the CIT using the GPT-4 API, to extract attack knowledge representations from textual descriptions.

\begin{figure*}[h!]
    \centering
\includegraphics[width=0.8\textwidth]{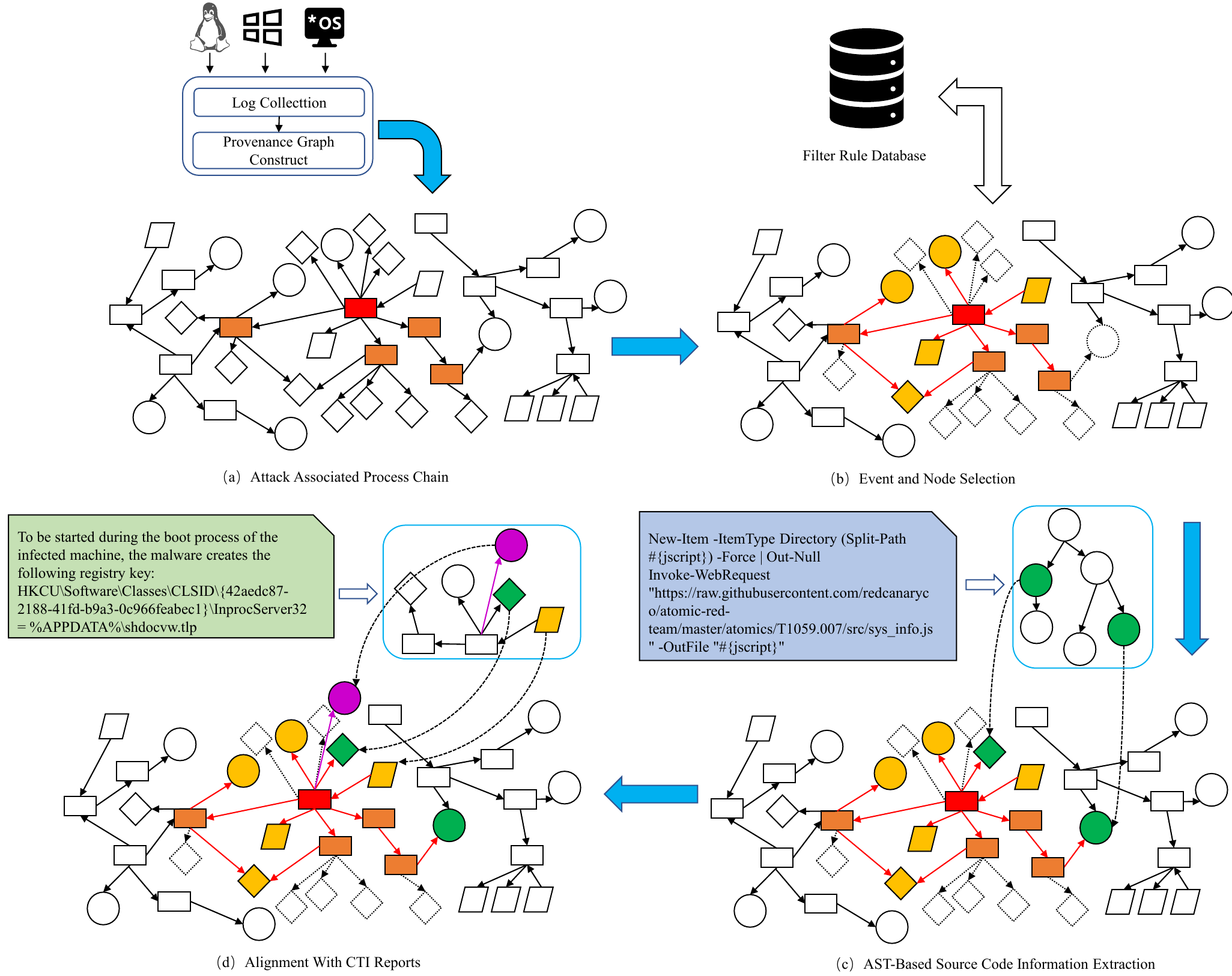}
        \caption{Attack Knowledge Graph Construction}
    \label{fig:mutilexample}
\end{figure*}

As presented in Figure~\ref{fig:mutilexample}, the local audit logs, which capture the real execution processes of different attacks, can help us obtain the most complete and realistic topology information about the attacks. 
Static code analysis can further refine the selection of nodes and their corresponding edges in the dynamic attack graph, making the attack graph representation more accurate.
Additionally, the extensive descriptions of the same attack technique across different threat reports can help us maximize the generalization of attack knowledge and supplement the knowledge of attack technique variants.

\subsection{Local attack technique graph from audit log}\label{SLOG}
The first step in building an atomic attack knowledge graph is to obtain the log event data dynamically executed by the attacker, then select the attack-related nodes and edges we need from it with a specific strategy, and finally complete the preliminary construction of the attack behavior knowledge graph. 
Note that we define an independent attack technique as an atomic attack.

Although it requires the execution of scripts or software, appropriate environment configurations as well as manual operations, and a corresponding log collector to obtain the data, this is the richest source for obtaining details about the execution of the attack technique and is necessary for constructing a fine-grained knowledge graph of the attack.
It is also a challenge to locate attack-related events from the whole system logs. A running system generates numerous entities and events continuously.
For a complete attack chain with different attack phases, accurately identifying each attack technique and its interrelationships is more complex.

In many cases, the starting process of an attack in a Provenance Graph will be connected to other nodes and edges that represent the different stages of the attack. 
However, our goal is to accurately identify the attack graph formed by the complete operation of a single atomic attack in a short time.
Since the execution of the attack is within our control, we have a lot of known information, such as the initial attack execution process ID, the execution period of the atomic attack, etc., which can effectively help us identify and construct attack subgraphs. We briefly present the entire process in algorithm \ref{alg:localknowledgegraph}, the specific steps are as follows.

\begin{algorithm}
    \caption{Construction of Local Attack Technique Graph from Audit Log}\label{alg:localknowledgegraph}
    \SetCommentSty{small}

    \KwData{System audit logs from attack execution}
    \KwResult{Atomic attack knowledge graph: $AttackGraph$}

    \tcp{Step 1: Automated Script Execution and System Event Collection}
    ConfigureAttackEnvironment()\;
    InitializeETWLogCollector()\;
    ExecuteAttackScript()\;
    $LogData \gets CollectEventLogs()$\;
    ExtractAttackDetails($LogData$)\;
    
    \tcp{Step 2: Process Association Chain Analysis}
    $InitialPID \gets GetAttackProcessID()$\;
    $ProcessChain \gets \{InitialPID\}$\;
    \For{each event $e$ in $LogData$}{
        \If{$e$ is associated with $InitialPID$}{
            UpdateProcessChain($e$, $ProcessChain$)\;
        }
    }
    RemoveCommonProcesses($ProcessChain$)\;

    \tcp{Step 3: Attack Event Selection}
    $SelectedEvents \gets \emptyset$\;
    \For{each event $e$ in $LogData$}{
        \If{$e$ is of relevant event type (e.g., ProcessStart, FileCreate, RegistrySet, etc.)}{
            \If{$e$ is not an irrelevant event type (e.g., ProcessEnd, ThreadEnd, FileioCreate, RegistryOpen)}{
                AddToSelectedEvents($e$)\;
            }
        }
    }

    \tcp{Step 4: Non-Attack Object Node Filtering}
    $Whitelist \gets CollectNonAttackNodes()$\;
    \For{each node $n \in SelectedEvents$}{
        \If{$n \notin Whitelist$}{
            AddToAttackGraph($n$)\;
        }
    }

    \tcp{Step 5: Edge and Label Aggregation}
    \For{each node $n \in AttackGraph$}{
        AggregateEdges($n$)\;
        SummarizeLabels($n$)\;
    }

    \Return $AttackGraph$\;
\end{algorithm}

\subsubsection{Automated Script Execution and System Event Collection}
we will combine automated tools and some manual operations to complete the attack environment configuration, and then realize script execution and system event data collection. We can fully automate the execution of the attack and collect the corresponding audit logs during execution.

First, we built a fully controllable system environment and implemented a lightweight log collection tool based on ETW. 
Next, we configured the framework of the Atomic Red Team. Based on this, we implement scripts to automate the patching of dependencies for the execution of atomic attacks.
Finally, we implemented scripts to automate the execution of atomic attacks and the collection of system audit log data during their execution. 

Based on this, we obtain information about the execution of the attack technique (e.g., execution period, attack-related process IDs, etc.), as well as complete system event data during execution.

\subsubsection{Process Association Chain Analysis}
Then, after getting the log event data, we began to screen the attack-related process chain. Since the environment in which the attack is performed is fully controllable, we can obtain the PID of the process where the attack started. Then, according to this Process ID (PID), we can find a series of associated process and thread chains that are subsequently associated, and these events constitute the Subject node in our attack graph.

We set a dynamically updated list of process IDs to selectively add specific processes associated with attack execution during the construction of the provenance graph to complete the construction of a complete attack process call chain.

The result is shown in Figure~\ref{fig:mutilexample}(a), Due to our prior understanding of the attack execution information, the integrity of the associated chain of the attack process that is screened out is very high. We can further filter out those common processes, such as hostname.exe, whoami.exe, etc., and add them in subsequent analysis.



\subsubsection{Attack Event Selection}
\begin{figure}[h!]
    \centering
    \includegraphics[width=0.4\textwidth]{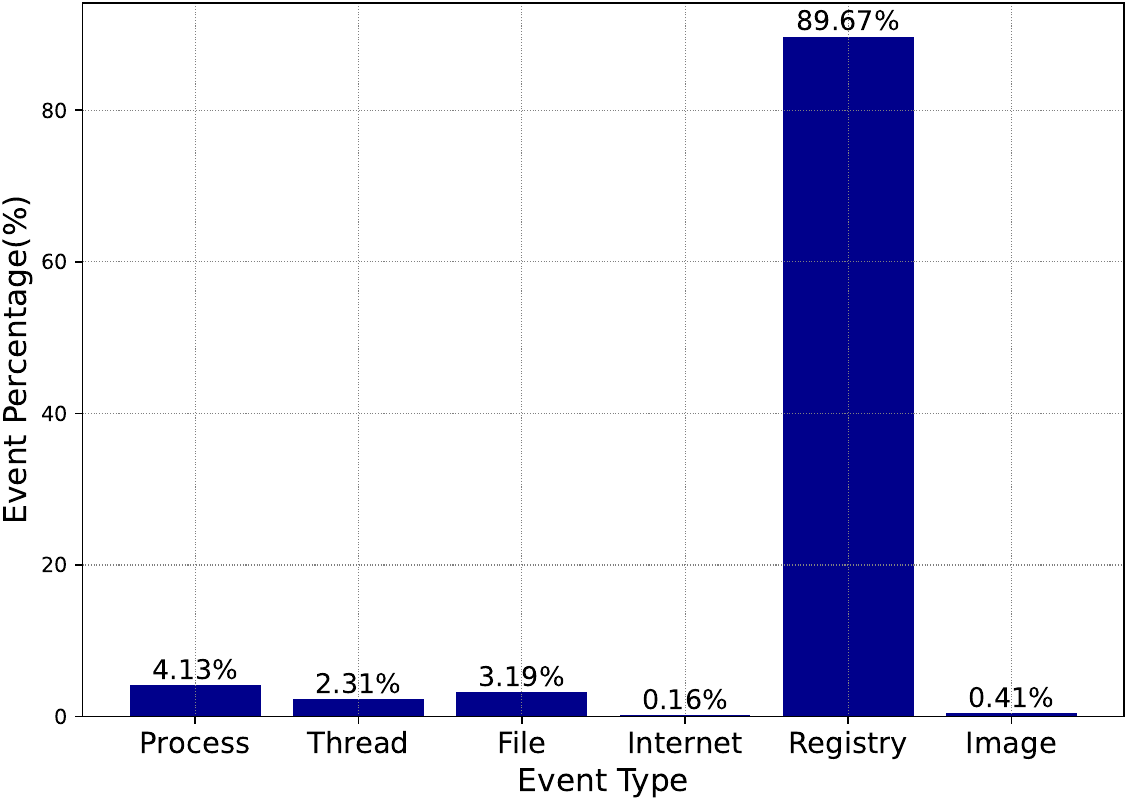}
    \caption{Statistics on the Percentage of Events in the System Audit Log}
    \label{fig:EventScale}
\end{figure}

According to our statistics on the scale of system events, as shown in Figure~\ref{fig:EventScale}, ETW\cite{ETW} can generate more than 6,000 system events per second under a low load state of the system,  among which registry events, file events, and process events account for a relatively high proportion.
When executing the attack technique, the system can generate more than 30,000 audit log events per second.

However, even if we have selected process and thread nodes in the previous step, there are still many events attached to these processes, and they are all regular system events that have nothing to do with the attack. 
As a result, our graph of atomic attack techniques can become very large, even involving hundreds of nodes and edges, which is far more than what is covered by a real attack execution.

In order to accurately represent an attack, we first have to make a choice about the event type. The event types we finally selected are shown in Table \ref{tab:Event Table}. We filtered a large number of useless events such as ProcessEnd, ThreadEnd, FileioCreate (for file handle acquisition), RegistryOpen, RegistryClose, and other irrelevant events, so as to more accurately represent atomic attacks. In addition, we only keep thread startup events under different processes, and the behavior of threads under the same process is merged into the parent process.

\begin{table}[h!]
\caption{Dependency Event Relationships based on Audit Log}
\label{tab:Event Table}
\scalebox{1.0}{
\begin{tabular}{lc}
\bottomrule[1.1pt]
\multicolumn{1}{l}{\textbf{Event Type}} & \textbf{Event Nmae}       \\ \hline
Process                                 & Start, DCStart           \\
Thread                                  & Start, DCStart\\
File                                    & Create, Read, Write, Rename           \\
Registry                                &  Query, Create, SetValue             \\
Internet                                & Receive, Send                \\
Image                                   & Load, DCStart                      \\ \bottomrule[1.1pt]
\end{tabular}}
\end{table}

\subsubsection{Non-Attack Object Node Filtering}
After the event filtering process, many reserved event object nodes are irrelevant to the attack. For example, when a process starts normally, it generates a large number of registry events, and file events, and generates and accesses temporary files. These specific registry or file paths are normal behavior of the process and have nothing to do with the attack.
We can also see from Figure~\ref{fig:EventScale} that there is a large percentage of such events.

In addition, there are also such things as random name files temporarily created by some programs for storing data or performing tasks, as well as registry paths associated with specific folders and their functions, and some specific system processes, etc., which need to be filtered.

Therefore, we need to filter the content of all reserved event nodes in order to retain the real attack-associated nodes. Since we execute the attack in a controlled environment, the security of the entire system can be guaranteed. 
Therefore, we can collect the non-malicious event node data of the system without performing any attack behavior and generate a whitelist for locating non-attack events. And, this data is utilized to filter the nodes when generating the attack graph to ensure the accuracy of the atomic attack graph.

\subsubsection{Edge and Label Aggregation}
Finally, after the screening of the above steps, we found that some atomic attack graphs still retain a large number of nodes. As a result of manual analysis, we found that most of these were a large number of operations of the same type, such as recursively accessing .doc files in the system directory, transferring to temporary files, and so on. We merge different operations on the same file and preserve all event-type labels. In addition, based on common file types and event types, we will merge edges and summarize node information to express attack knowledge more concisely and universally.

\subsection{Attack Script Static Analysis}\label{SAST}


Although a relatively complete attack subgraph structure can be obtained through the audit log, however, due to various screening and filtering methods in order to obtain an accurate and concise attack subgraph, we may lose individual nodes of the subgraph in some specific attack techniques.

We realized that missing nodes and edges in dynamic graphs could be completed from the most informative static code analysis. We use the Abstract Syntax Tree (AST) to parse the static code, obtain the nodes we need from the AST through a specific algorithm, and then complete the supplement to the dynamic execution subgraph.

Specifically, we can extract the attack-related entity information from the AST and match it with the corresponding data in the audit logs. This helps us to select the node and the events related to the entity, and then add them to the dynamic attack graph, which makes the representation of the attack more accurate. We have shared the entire processing procedure in algorithm \ref{alg:staticanalysis}.

\begin{algorithm}
    \caption{Construction of Attack Technique Graph using Static Analysis}\label{alg:staticanalysis}
    \SetCommentSty{small}

    \KwData{Attack script source code}
    \KwResult{Enhanced attack knowledge graph: $AttackGraph$}

    \tcp{Step 1: Parse AST from Source Code}
    $AST \gets GenerateAST(SourceCode)$\;
    $StaticNodes \gets \emptyset$\;
    $SupplementGraph \gets \emptyset$\;

    \tcp{Step 2: AST Node Selection}
    \For{each node $n$ in $AST$}{
        \If{$n$ is of type StringConstantExpressionAst}{
            SaveNodeValue($n$, $StaticNodes$)\;
        }
        \If{$n$ is of type VariableExpressionAst}{
            $VariableMap \gets ExtractVariableMapping(n)$\;
            SaveToMap($VariableMap$)\;
        }
        \If{$n$ is of type ExpandableStringExpressionAst}{
            $ExpandedValue \gets SimulateAndExpand($n$)$\;
            AddToStaticNodes($ExpandedValue$)\;
        }
        \If{$n$ is a CommandAst node}{
            $Command \gets ExtractFirstStringConstant(n)$\;
            SaveCommand($Command$, $StaticNodes$)\;
        }
    }

    \tcp{Step 3: Node Information Analysis and Mapping}
    \For{each node $s$ in $StaticNodes$}{
        \If{$s$ is relevant to registry, file, or process entity}{
            MatchWithAuditLog($s$)\;
            AddToSupplementGraph($s$)\;
        }
    }

    \tcp{Step 4: Construct Enhanced Attack Graph}
    MergeGraphs($AttackGraph$, $SupplementGraph$)\;

    \Return $AttackGraph$\;
\end{algorithm}


\subsubsection{AST Node Selection}
\begin{figure*}[h!]
	\centering
	\subfigure[An example graph of an abstract syntax tree]{
  \includegraphics[width=0.45\textwidth]{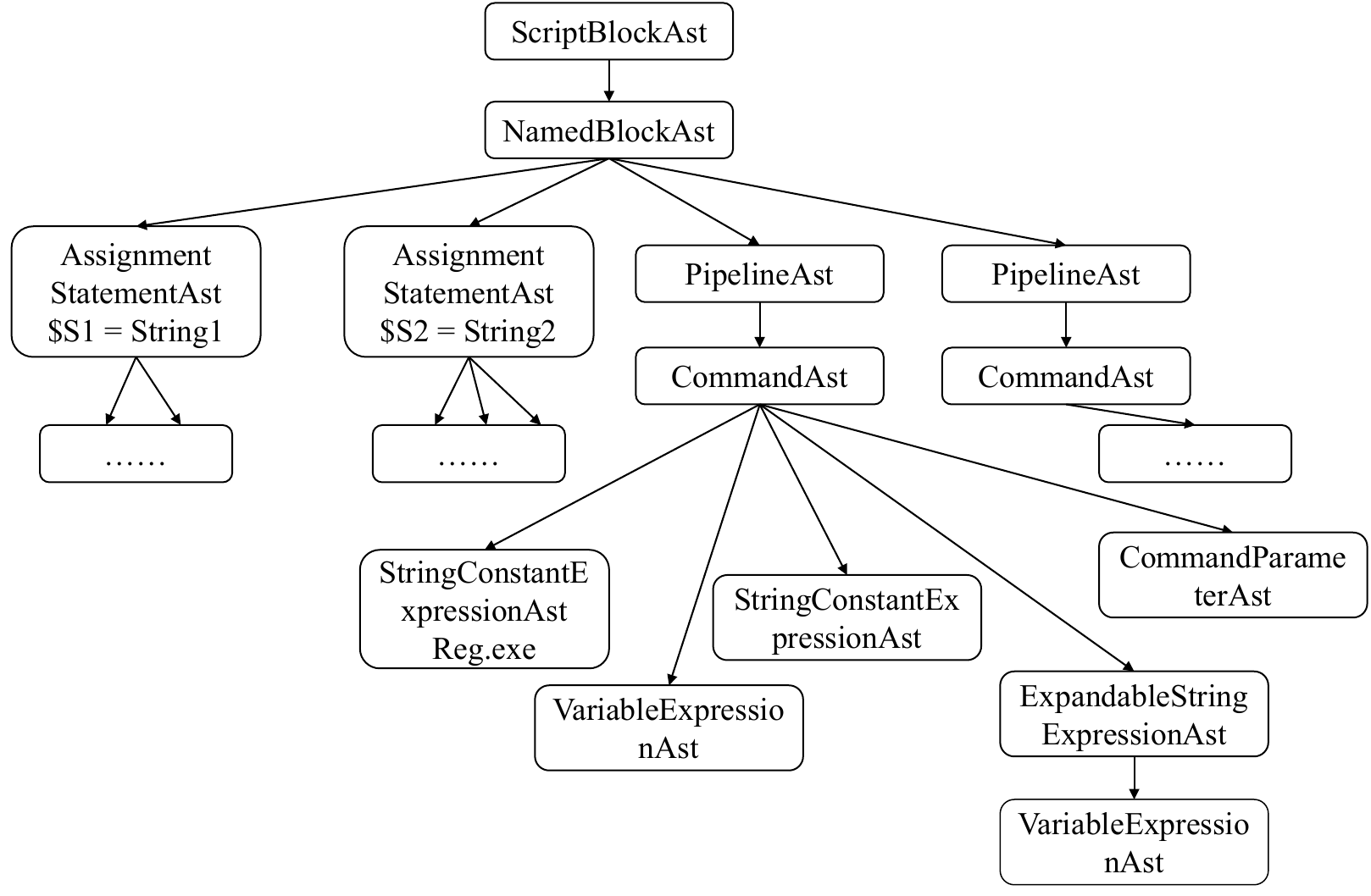}
    \label{fig:AST1}
	}	
	\subfigure[The information we select from the abstract syntax tree.]{
    \includegraphics[width=0.45\textwidth]{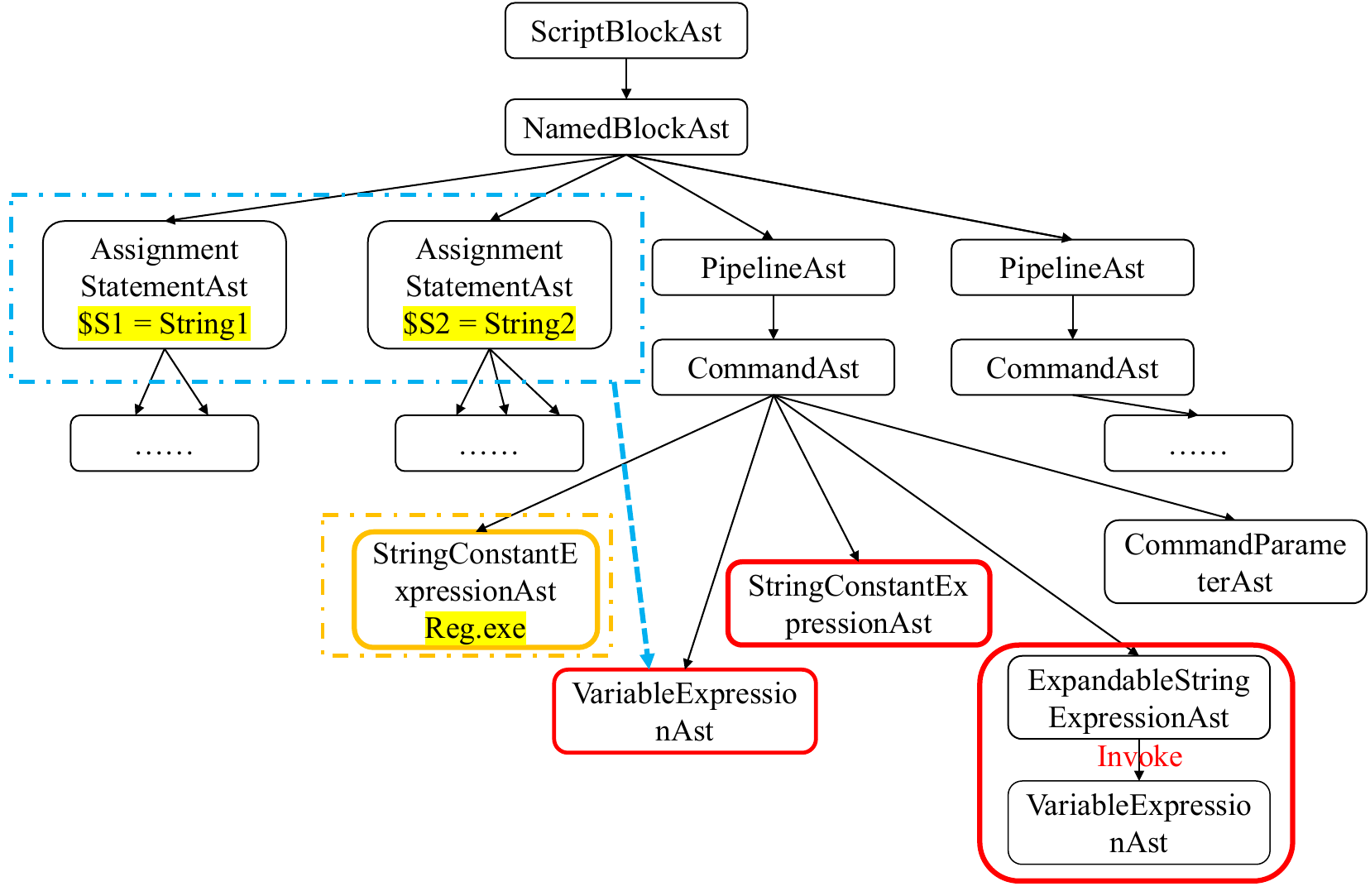}
    \label{fig:AST2}
	}
    \caption{Static Analysis of Attack Scripts Based on Abstract Syntax Tree.}
	\label{Fig: serveroverhead3image}
\end{figure*}

Figure~\ref{fig:AST1} is an example of AST, which is a tree-shaped hierarchical structure composed of several nodes, such as ScriptBlockAst,
PipelineAst, StringConstantExpressionAst, etc.

Since we have identified the process association chain, we now need to find the object node pointed to by the process through the AST. Therefore, we need to retrieve the relevant entity names such as registry and file names in the AST and save them to determine whether we need to add edges to related object nodes during the dynamic construction process. 

We mainly focus on the contents of the following nodes, namely StringConstantExpressionAst, VariableExpressionAst, and ExpandableStringExpressionAst. The node types that assist us in node content processing are CommandAst and AssignmentStatementAst.


\subsubsection{Node Information Analysis and Processing}
As shown in Figure~\ref{fig:AST2}, first, we save StringConstantExpressionAst, which is a static variable, so we can directly save its value, followed by VariableExpressionAst, which is a dynamic variable name. We need to save AssignmentStatementAst in the syntax tree in the form of a map, and replace it here, so as to achieve and audit The content of the log is unified. Next is ExpandableStringExpressionAst, this type of node requires us to do dynamic simulation to restore its full name in the system. Finally, we found that the first StringConstantExpressionAst node under the CommandAst node generally stores the relevant command or process name, and we also save it to assist in the construction of our process chain.

We conduct static analysis to filter out a series of entity nodes related to processes, files, and registries. These nodes are then used to recover their corresponding events and relationships when we construct the dynamic graph with the audit log.


\subsection{CTI Reports Analyis}\label{SCTI}
Through dynamic audit logs and static source code analysis, we have ensured that we can obtain a complete attack subgraph of an atomic attack technology. However, such a limitation is that we cannot obtain a relatively complete implementation of a certain attack method. At the same time, threat intelligence just provides rich and high-value information for guiding log analysis of real attack behaviors, which can help us improve the construction of attack subgraphs.

For example, in the persistent attack technique T1547, one of the ways is to modify the registry to realize the self-starting of malware. There are many registry directories that can be used to modify, such as:
\begin{list}
{\labelitemi}{\leftmargin=1.5em}

    \item \textit{HKEY\_CURRENT\_USER\textbackslash Software\textbackslash...\textbackslash CurrentVersion\textbackslash Explorer}
    \item \textit{HKEY\_LOCAL\_MACHINE\textbackslash Software\textbackslash ...\textbackslash RunServicesOnce}
    \item \textit{HKEY\_CURRENT\_USER\textbackslash Software\textbackslash...\textbackslash Policies\textbackslash Explorer\textbackslash Run}
    \item \textit{HKEY\_LOCAL\_MACHINE\textbackslash System\textbackslash...\textbackslash Control\textbackslash Session Manager}
    
\end{list}

In addition to this, many more registry directories can be used to help attackers achieve persistence. However, for a specific executable file and attack code, may only contain a specific registry directory to complete the attack. If we want to fully represent this type of attack, for example by modifying the registry to achieve persistence, then Cyber Threat Intelligence (CTI) is undoubtedly the best choice. There are tens of thousands of reports available on the Internet, providing a large number of different implementations of the same technology with nuances, which can be used by us to improve the attack subgraph generated above, so as to make our subgraph representation more generalized and more accurate.

Through algorithm \ref{alg:ctianalysis}, we briefly describe the entire procedure.
We will perform the following operation to manipulate the attack knowledge from the CTI reports:
\begin{list}
{\labelitemi}{\leftmargin=1.5em}
    \item Technique procedure example crawled from MITER ATT\&CK contains descriptions of the same attack technique from different reports.
    \item Based on the GPT4 API and a well-designed prompt, we can parse the report text and get the entities, entity types, and inter-entity relationships.
    \item For multiple reports on the same technique, proper alignment and merging of the knowledge graph allows us to obtain a unified representation of a specific attack technique.
\end{list}

\begin{algorithm}
    \caption{CTI Reports Analysis for Attack Knowledge Graph Enhancement}\label{alg:ctianalysis}
    \SetCommentSty{small}

    \KwData{CTI reports on attack techniques}
    \KwResult{Enhanced attack knowledge graph: $AttackGraph$}

    \tcp{Step 1: Retrieve Technique Descriptions from MITRE ATT\&CK}
    $TechniqueDescriptions \gets CrawlMITREATTACK()$\;
    $TechniqueGraphs \gets \emptyset$\;
    
    \For{each description $d$ in $TechniqueDescriptions$}{
        $TechGraph \gets GenerateGraphFromDescription(d)$\;
        AddToTechniqueGraphs($TechGraph$, $TechniqueGraphs$)\;
    }

    \tcp{Step 2: Parse CTI Reports for Entity and Relationship Extraction}
    \For{each report $r$ in $CTIReports$}{
        $Entities, Relationships \gets ParseReportWithGPT4(r)$\;
        ExtractIOCInfo($Entities$)\;
        GeneratePartialGraph($Entities$, $Relationships$)\;
    }

    \tcp{Step 3: Aggregate Technique Graphs Across Reports}
    \For{each technique $t$ in $TechniqueGraphs$}{
        $UnifiedGraph \gets InitializeUnifiedGraph(t)$\;
        
        \For{each partial graph $g$ for technique $t$}{
            AlignAndMergeGraph($g$, $UnifiedGraph$)\;
        }
        
        $AttackGraph \gets MergeIntoAttackGraph(UnifiedGraph, AttackGraph)$\;
    }

    \Return $AttackGraph$\;
\end{algorithm}

While CTI reports provide detailed contexts about an attack (e.g., the sequence of adversarial actions), manually extracting attack-relevant information from uninstructed texts is labor-intensive and error-prone, hindering CTI's practice applications.
Based on the observation that IoCs often share similar context terms, iACE~\cite{Liao} presents a graph mining technique to collect IoCs available in tens of thousands of security articles.
ChainSmith~\cite{Zhu2018} further makes use of neural networks to classify IoCs into different attack campaign stages (e.g., baiting and C\&C).
TTPDrill~\cite{Husari2017} derives threat actions from Symantec reports and maps them to attack patterns (e.g., techniques in MITRE ATT\&CK~\cite{mitre}) pre-defined as ontology.
EXTRACTOR~\cite{Satvat2021} and ThreatRaptor~\cite{gao} customize NLP techniques to model attack behaviors in texts as provenance graphs. AttacKG~\cite{li2022attackg} adapting NLP models based on limited data to recognize entities as well as their relationships.

However, \SysName differs from these works in that it parses CIT reports and constructs knowledge graphs by using LLM, which has a large amount of security domain knowledge data and strong contextual understanding and knowledge inference capabilities~\cite{touvron2023llama, brown2020language}.

\subsubsection{Attack Technique Procedure}
First, we need general textual descriptions of individual attack techniques. In this paper, with the threat report descriptions on attack technique procedure provided in MITRE ATT\&CK, we construct multiple graph-structured technique descriptions for each technique.

Specifically, we use the crawler algorithm to obtain attack technology cases from MITRE ATT\&CK, and the attack technology cases describe the details of each attack technology.
Based on the behavioral descriptions of each attack technique described in the crawled reports, we construct multiple attack knowledge graphs about the same technique using LLMs, which are then aggregated by algorithms.

\subsubsection{CTI Reports Parser}
Well-written CTI reports include detailed technical descriptions of how attack-related entities interact to accomplish adversarial objectives
in attack campaigns. Despite the rich information, it is challenging to accurately
extract attack behaviors from CTI reports written in natural language. 

To solve this problem, we design a new CTI report parsing scheme based on the existing one, which sequentially performs the identification of entities and entity types, as well as the judgment of inter-entity relationships.

Specifically, we build a parser to parse the CTI report based on the API provided by the large language model GPT-4, design a prompt to extract the entities and entity dependencies related to the attack and use regular expressions to recognize fine-grained IOC information to construct, generate, and simplify the attack graph. Eventually, we were able to develop a concise and clear attack graph describing the various attack behaviors reported.

\subsubsection{Attack Knowledge Graph Generation}
A single attack report often only describes an implementation of several attack techniques, lacking a summary of its overall implementation. We therefore compensate for this by aggregating threat intelligence across reports.

Now, we utilized a graph alignment algorithm to match technical templates in attack graphs.

We need to merge the multiple attack knowledge graphs about the same technique extracted from the analysis of the above example technique descriptions. 
We designed graph alignment algorithms to identify the attack techniques and find the corresponding nodes and edges in the attack graphs and as a way to merge the attack graphs of techniques extracted from multiple CTI reports.

After finding the corresponding nodes and edges, we can utilize the information contained in both to complement each other. In this case, the single-technique graph contains the execution process of a certain technique, and the multi-technique attack graph from CTI reports contains real attack cases consisting of multiple techniques and their realization details.


Clearly, we can accurately align and refine entities in CTI reports and technology templates, building a knowledge summary graph of attack-related knowledge across multiple reports. We will introduce in \S \ref{S5} the way to merge multiple descriptive attack resolution graphs from multiple CIT reports about the same attack technique.

\section{Knowledge Graph Construction}\label{S5}
In this section, we introduce two parts: 1) how to do aggregation of multiple knowledge graphs about the same attack technique from a single data source. 2) how to do a fusion of knowledge graphs from different sources to generate a unified and complete attack knowledge graph representation.




\subsection{Merging of Technical Graphs from the Same Resource}\label{S51merghsingle}
Based on the above design, we obtained multiple technical graphs from audit logs and CTI reports respectively. 
Before getting the final cross-source technical representation, we need to merge multiple technical diagrams from the same source into one complete representation, respectively. In other words, multiple same-source graphs corresponding to the same technique are merged.
\begin{algorithm}[h]
    \caption{Merging of Technical Graphs from the Same Source}\label{alg:merginggraphs}
    \SetCommentSty{small}
    
    \KwData{Multiple technical graphs from audit logs or CTI reports: $Graphs_{source}$}
    \KwResult{Merged technical graph: $MergedGraph$}
    
    $MergedGraph \gets \emptyset$\;
    \tcp{Step 1: Merge attacker nodes from different attack graphs}
    \For{each graph $G \in Graphs_{source}$}{
        $AttackerNodes \gets \{n \mid n \text{ is an attacker node in } G\}$\;
        $MergedGraph \gets MergedGraph \cup AttackerNodes$\;
    }

    \tcp{Step 2: Merge nodes at the same hierarchical level}
    \For{each level $l$ in $Graphs_{source}$}{
        \For{each node type $t$ at level $l$}{
            \uIf{$t$ is Process type}{
                $SameLevelNodes \gets \{n \mid n \text{ is at level } l \text{ with same process name in } Graphs_{source}\}$\;
            }
            \Else{
                $SameLevelNodes \gets \{n \mid n \text{ is at level } l \text{ with type } t \text{ in } Graphs_{source}\}$\;
            }
            $MergedGraph \gets MergedGraph \cup Merge(SameLevelNodes)$\;
        }
    }

    \tcp{Step 3: Merge nodes across layers with identical attributes}
    \For{each attribute $attr$ in $Graphs_{source}$}{
        $NodesWithAttr \gets \{n \mid n \text{ has attribute } attr \text{ in } Graphs_{source}\}$\;
        $MergedGraph \gets MergedGraph \cup MergeAcrossLayers(NodesWithAttr)$\;
    }

    \tcp{Step 4: Merge leaf nodes based on content similarity (e.g., images, files)}
    \For{each leaf node $n$ in $Graphs_{source}$}{
        $SimilarNodes \gets FindSimilarNodes(n)$\;
        $MergedGraph \gets MergedGraph \cup ClusterAndMerge(SimilarNodes)$\;
    }

    \tcp{Step 5: Preserve edges and add extra attributes to merged nodes}
    \For{each merged node $m \in MergedGraph$}{
        RetainEdges($m$)\;
        AddExtraAttributes($m$)\;
    }

    \tcp{Step 6: Generalize node information for universal representation}
    GeneralizePaths($MergedGraph$)\;
    ApplyWildcardSubstitutions($MergedGraph$)\;

    \Return $MergedGraph$\;
\end{algorithm}

First, we merge the attacker nodes from the different attack graphs generated by the logs to make a connected graph.
Then, we do the merging strategy for its nodes of the same level (level refers to the hierarchy of each associated entity from the starting node, directly connected as level-1, subsequent as level-2, etc.). The starting node is the first node in the GML format description, i.e., the attacker node (id attribute is 0)).

For Process type nodes, we merge nodes at the same level with the same process name.
For the rest of the nodes, we merge the nodes at the same level and of the same type.

The idea of merging at the same level is easy to understand, similar nodes follow similar calling relationships, share the same calling logic and are more likely to appear at the same level of the calling chain, which is confirmed by our experimental results. 
This means that the contents of the nodes corresponding to the same entity are all recorded among the additional attributes, and the corresponding nodes are merged in the construction of the attack graph.

In addition, for nodes with the same attribute content, we will also merge them across layers so that the technical graph is more concise without redundant nodes.

Then, we also merge the object nodes such as images, files, and registries located in the leaf nodes of the technical graph by extracting the common prefix. We do node clustering and merging based on the similarity of node contents, which makes the attack graph representation more generalized.

Note that for nodes that have been merged, we first ensure that the relationships of the edges previously attached to that node remain unchanged. Then, we add an extra attribute to the merged node, which includes all the label attributes for the merged node.

Finally, we will generalize and de-duplicate some node information to make its representation more universal. For example, we will represent
\texttt{C\textbackslash:\textbackslash\textbackslash Users\textbackslash\textbackslash Alice\textbackslash\textbackslash AppData\textbackslash\textbackslash Local}
as \texttt{C\textbackslash:\textbackslash\textbackslash Users\textbackslash\textbackslash.*\textbackslash\textbackslash AppData\textbackslash\textbackslash Local}, and {HKEY\_LOCAL\_MACHINE\textbackslash System\textbackslash .*} to match specific registry sets and so on. In addition, we will also do wildcard substitutions for some temporary system file names and so on.


In the end, for the same attack technique, we get one merged technique diagram from the audit log and CIT report respectively. We present the complete algorithmic processing logic in algorithm \ref{alg:merginggraphs}.
\subsection{Construction of technique-level attack knowledge graph from multi-source}\label{S52}
With the above analysis, we get a separate technical graph from each of the different sources of attack information.

For now, we need to merge the technology graphs across sources to get a unified and complete attack knowledge graph for each technology. We aim to merge different graphs corresponding to the same technique across sources.
We find the eligible technique graphs based on the technique number. Then, load them and represent them as a unified technology graph following the merging strategy.

For all nodes of the technical graph under different resources, we create multiple lists storing the breadth-first search order of the nodes in each graph.
Then, we define the graphs that are produced based on both the audit log and static code as the base graph. The ones based on CIT reports are defined as the additional graph. Lastly, we create a new graph to be the final results graph.

For each node in the additional graph, we iterate according to the node queue order. We check whether there is a matching node of the same type in the basic graph and its content similarity with the corresponding type of node, respectively, to determine whether it matches and needs to be merged.
If a match is found, we update the list of nodes in the new graph, remove the merged node from the list of nodes in the base graph and the additional graph, and update the data of the merged node and its corresponding edges to the new graph. 
If no match is found in the base graph, we add the node to the new graph and set an edge connecting the node to its parent.
This parent node is the upper-level node corresponding to this node in the additional graph, which must exist in the current new graph due to the traversal according to the breadth-first search.

Lastly, since the merged node will contain the data of all the nodes that were merged, we will perform deduplication and generalization to ensure the universality of the final technique knowledge graph.

\begin{algorithm}
    \caption{Construction of Technique-Level Attack Knowledge Graph from Multiple Sources}\label{alg:constructiongraph}
    \SetCommentSty{small}

    \KwData{Technical graphs from different sources: $TechGraphs_{sources}$}
    \KwResult{Unified attack knowledge graph: $UnifiedGraph$}

    \tcp{Step 1: Initialize base and additional graphs}
    $BaseGraph \gets \text{graph from audit log or static code}$\;
    $AdditionalGraph \gets \text{graph from CTI reports}$\;
    $UnifiedGraph \gets \emptyset$\;
    
    \tcp{Step 2: Select and load eligible technique graphs by technique number}
    \For{each technique number $t$ in $TechGraphs_{sources}$}{
        $GraphsToMerge \gets \{G \mid G \text{ corresponds to technique } t \}$\;
        LoadGraphs($GraphsToMerge$)\;
    }

    \tcp{Step 3: Prepare BFS order lists for each source graph}
    $BFSOrder_{Base} \gets BFS(BaseGraph)$\;
    $BFSOrder_{Additional} \gets BFS(AdditionalGraph)$\;

    \tcp{Step 4: Merge nodes from additional graph into base graph}
    \For{each node $n_{add} \in BFSOrder_{Additional}$}{
        $MatchingNode \gets FindMatchingNode(n_{add}, BFSOrder_{Base})$\;
        
        \uIf{$MatchingNode$ exists and ContentSimilarity($n_{add}$, $MatchingNode$) is high}{
            UpdateNodeList($UnifiedGraph$, $MatchingNode$)\;
            RemoveNode($n_{add}$, $BFSOrder_{Base}$)\;
            UpdateEdges($UnifiedGraph$, $MatchingNode$)\;
        }
        \Else{
            AddNode($UnifiedGraph$, $n_{add}$)\;
            $ParentNode \gets FindParentNode(n_{add}, UnifiedGraph)$\;
            SetEdge($ParentNode$, $n_{add}$, $UnifiedGraph$)\;
        }
    }

    \tcp{Step 5: Perform deduplication and generalization for universal representation}
    DeduplicateNodes($UnifiedGraph$)\;
    GeneralizeNodeInfo($UnifiedGraph$)\;

    \Return $UnifiedGraph$\;
\end{algorithm}







%% file: 7_Evaluation.tex
\section{Evaluation}\label{Evaluation}
\subsection{Evaluation Setup}
To evaluate the \SysName, we utilized the source code and executable frameworks of the 1015 attack procedures of the 282 attack techniques crawled from the atomic red team as the source of the base dataset.
Then, based on the technique numbers corresponding to these attack techniques in the atomic red team, we crawled all the threat report procedure descriptions corresponding to a single attack technique from the technique details page in MITRE ATT\&CK, respectively, totaling 9006 attack report descriptions. Statistics for the 1,015 attack techniques involved are provided in Table \ref{StatisticaldataForART}.

\begin{table*}[h!]  
\caption{Statistical Overview of the Atomic Attack Techniques Dataset} 
\label{StatisticaldataForART}
\begin{tabular}{cccc}  
\bottomrule[1.1pt]  
\multicolumn{1}{c}{Tactic} & Technique & Sub-Technique & Description in MITRE ATT\&CK \\ \hline  
Defense-Evasion            & 71        & 323           & avoid being detected.             \\  
Privilege-Escalation       & 43        & 122           & gain higher-level permissions.             \\  
Execution                  & 11        & 72            & run malicious code.             \\  
Persistence                & 46        & 104           & maintain attack foothold.             \\  
Collection                 & 14        & 26            & gather data of interest to their goal.             \\  
Lateral-Movement           & 9         & 21            & move through your environment.             \\  
Credential-Access          & 30        & 109           & steal account names and passwords.             \\  
Discovery                  & 27        & 147           & figure out your environment.             \\  
Command-and-Control        & 11        & 49            & communicate with compromised systems to control them.             \\  
Reconnaissance             & 1         & 1             & gather information they can use to plan future operations.             \\  
Impact                     & 7         & 22            & manipulate, interrupt, or destroy your systems and data.             \\  
Initial-Access             & 6         & 10            & get into your network.             \\  
Exfiltration               & 6         & 9             & steal data from your network.            \\ \hline  
Overall                    & 282       & 1015          & -             \\ \bottomrule[1.1pt]  
\end{tabular}  
\label{tab:atomic attack graph}  
\vspace{-0.0in}  
\end{table*}  

We have implemented an efficient data collection system based on the native Windows ETW in C++ to ensure efficient and high-fidelity collection of system logs during the execution of the attack technique.
We also implemented AST parsing and node analysis of the attack code based on C\# to achieve static analysis and attack information acquisition.
In addition, we utilize the GPT-4 API to complete the text parsing and processing of attack technique descriptions in CIT reports.

Also, we manually annotate the entities, entity types, and relationships of a portion of the attack techniques in the audit logs and reports, respectively. This is used as our ground truth for manual evaluation to show the accuracy of our method.


\subsection{Evaluation Results}
We evaluate \SysName by answering the following questions.
\begin{list}
{\labelitemi}{\leftmargin=1.5em}

    \item RQ1: How accurate is \SysName in extracting attack technique information from various sources?
    \item RQ2: How effective is \SysName in aggregating technical level knowledge?
    \item RQ3: How effective is the attack knowledge graph constructed by \SysName at the technique level?
    \item RQ4: How accurate is \SysName in attack investigation?
    
\end{list}

To answer RQ1, we manually analyzed the execution logs of the ten attack techniques that appeared with high frequency and manually labeled the attack techniques from the different phases involved in eleven reports of real-world APT activity to compare with the results generated by \SysName.
To answer RQ2, we statistically represent all the attack knowledge graphs from the atomic red team and CTI reports, and show some of the before and after aggregation results to demonstrate the effectiveness of our aggregation algorithm.
To answer RQ3, we compare \SysName with AttacKG\cite{li2022attackg} which is also at the Technique-aware level and has a graph structure.
We demonstrate the effectiveness of our method in terms of the information content of the nodes and edges in the graph, and in comparison with the manually generated ground-truth, respectively.
Finally, we give a case study to illustrate how \SysName can benefit downstream security tasks.

\subsubsection{Accuracy of information on attack techniques extracted from various sources}

\begin{table*}[h!]
\caption{Accuracy of 10 attack technique graph extraction from the dynamic audit log with static code assistance. (Columns 3-6 show nodes and edges for manual extraction of attack techniques, and \textcolor{red}{+false
\_positive} (\textcolor{blue}{-false\_negative}) for \SysName-generated graphs. Rows 13-15 represent the overall Precision, Recall, and F1-Score.)}  
\label{logsingle}
\begin{tabular}{@{}l|c|cccc@{}}
\toprule
\multirow{2}{*}{Technique}  & \multirow{2}{*}{Description}  & \multicolumn{2}{c}{Node}                            & \multicolumn{2}{c}{Edge} \\ \cmidrule(l){3-6} 
                           &                             & Manual                & \SysName                        & Manual                & \SysName   \\ \midrule
T1547.001-1          & Registry Run Keys / Startup Folder     & 6                     & \multicolumn{1}{c|}{\textcolor{red}{+0} (\textcolor{blue}{-0})} & 5                     & \textcolor{red}{+0} (\textcolor{blue}{-0}) \\
T1003.003-6           & OS Credential Dumping: NTDS    & 5                      & \multicolumn{1}{c|}{\textcolor{red}{+1} (\textcolor{blue}{-1})}       & 4                       & {\textcolor{red}{+0} (\textcolor{blue}{-2})}   \\
T1562.001-30                           & Kill the event log services for stealth via function of WinPwn                  & 5                      & \multicolumn{1}{c|}{\textcolor{red}{+0} (\textcolor{blue}{-0})}       & 4                       & {\textcolor{red}{+0} (\textcolor{blue}{-0})}       \\ 

T1105-9                           & Utilize BITSAdmin to schedule jobs for downloading malware                  & 7                      & \multicolumn{1}{c|}{\textcolor{red}{+0} (\textcolor{blue}{-1})}       & 7                       & {\textcolor{red}{+1} (\textcolor{blue}{-2})}       \\ 

T1112-14                           & Executes signed PubPrn script with options to execute the payload                & 7                      & \multicolumn{1}{c|}{\textcolor{red}{+0} (\textcolor{blue}{-0})}       & 6                       & {\textcolor{red}{+1} (\textcolor{blue}{-1})}       \\ 

T1216.001-1                           & Modify the configuration to disable shutdown button.                  & 9                      & \multicolumn{1}{c|}{\textcolor{red}{+0} (\textcolor{blue}{-0})}       & 8                       & {\textcolor{red}{+0} (\textcolor{blue}{-0})}       \\

T1078.001-1                           & Default Guest activated, added to Admin and RDP group                  & 18                      & \multicolumn{1}{c|}{\textcolor{red}{+3} (\textcolor{blue}{-2})}       & 24                       & {\textcolor{red}{+3} (\textcolor{blue}{-2})}       \\

T1485-1                           & Overwrite file with SysInternals SDelete                 & 6                      & \multicolumn{1}{c|}{\textcolor{red}{+1} (\textcolor{blue}{-0})}       & 5                       & {\textcolor{red}{+1} (\textcolor{blue}{-0})}       \\

T1490-9                           & Disable System Restore Through Registry                 & 12                      & \multicolumn{1}{c|}{\textcolor{red}{+0} (\textcolor{blue}{-1})}       & 14                       & {\textcolor{red}{+1} (\textcolor{blue}{-1})}       \\

T1037.001-1                           & Add registry value to run the batch script created in temp directory               & 6                      & \multicolumn{1}{c|}{\textcolor{red}{+0} (\textcolor{blue}{-0})}       & 5                       & {\textcolor{red}{+0} (\textcolor{blue}{-0})}       \\

\midrule
Overall Precious                    &  &  \multicolumn{1}{c}{1.000} & \multicolumn{1}{c|}{0.938} & \multicolumn{1}{c}{1.000} & 0.914        \\
Overall Recall                    & - &  \multicolumn{1}{c}{1.000} & \multicolumn{1}{c|}{0.938} & \multicolumn{1}{c}{1.000} & 0.902        \\ 
Overall F-1 Score                    &  &  \multicolumn{1}{c}{1.000} & \multicolumn{1}{c|}{0.938} & \multicolumn{1}{c}{1.000} & 0.908        \\ 
\bottomrule
\end{tabular}
\label{tab:RQ1LogResult}  
\end{table*}

\begin{table*}[h!]
\caption{Comparison of Accuracy in Extracting Attack Technique Graphs from 11 CTI Reports. (Columns 2-5 list the manually generated counts of ground truths and \textcolor{red}{+false\_positive} (\textcolor{blue}{-false\_negative}) in extracting entity nodes and entity relationships. Columns 6-10 list the counts of manually generated ground truths and \textcolor{purple}{+false\_positive} (\textcolor{orange}{-false\_negative}) in extracting and determining the type of entity nodes associated with the attack. Rows 14-16 present the overall Precision, Recall, and F1-Score.)}
\label{singleCIT}

\begin{tabular}{@{}l|ccccccccc@{}}
\toprule
\multirow{2}{*}{Technique \& Procedure} & \multicolumn{2}{c}{Node}       & \multicolumn{2}{c}{Edge}       & \multicolumn{5}{c}{Node Type}                    \\ \cmidrule(l){2-10} 
                           & Manual & \SysName                 & Manual & \SysName                 & Process & File & Registry & Image & Network \\ \midrule
T1547-BoxCaon                           & 2       & \multicolumn{1}{c|}{\textcolor{red}{+0} (\textcolor{blue}{-0})} &    1    & \multicolumn{1}{c|}{\textcolor{red}{+0} (\textcolor{blue}{-0})} & \textcolor{purple}{+0} (\textcolor{orange}{-0})        &  -    & \textcolor{purple}{+0} (\textcolor{orange}{-0})         &  -     &   -      \\
T1218.008-Bumblebee                           & 4       & \multicolumn{1}{c|}{\textcolor{red}{+0} (\textcolor{blue}{-0})} &  3      & \multicolumn{1}{c|}{\textcolor{red}{+0} (\textcolor{blue}{-0})} &  \textcolor{purple}{+0} (\textcolor{orange}{-0})       & \textcolor{purple}{+0} (\textcolor{orange}{-0})     &     -     &    -   &  \textcolor{purple}{+0} (\textcolor{orange}{-0})       \\

    T1036.003-CozyCar                       &  3      & \multicolumn{1}{c|}{\textcolor{red}{+1} (\textcolor{blue}{-0})} &    3    & \multicolumn{1}{c|}{\textcolor{red}{+1} (\textcolor{blue}{-1})} &  \textcolor{purple}{+0} (\textcolor{orange}{-1})       &  \textcolor{purple}{+2} (\textcolor{orange}{-0})    &     -     &     \textcolor{purple}{+0} (\textcolor{orange}{-0})  &    -     \\ 

    T1115-Attor                       &   5     & \multicolumn{1}{c|}{\textcolor{red}{+1} (\textcolor{blue}{-0})} &     8   &  \multicolumn{1}{c|}{\textcolor{red}{+1} (\textcolor{blue}{-0})} &    \textcolor{purple}{+0} (\textcolor{orange}{-0})     &  \textcolor{purple}{+1} (\textcolor{orange}{-0})    &     -     &    -   &    -     \\ 
    T1490-BlackCat                     &  7      & \multicolumn{1}{c|}{\textcolor{red}{+1} (\textcolor{blue}{-1})} &   6     & \multicolumn{1}{c|}{\textcolor{red}{+1} (\textcolor{blue}{-1})} &      \textcolor{purple}{+0} (\textcolor{orange}{-0})   &   \textcolor{purple}{+0} (\textcolor{orange}{-0})   &     -     &   -    &   -      \\ 
    T1489-Industroyer                       &   3     & \multicolumn{1}{c|}{\textcolor{red}{+2} (\textcolor{blue}{-0})} &   2     & \multicolumn{1}{c|}{\textcolor{red}{+3} (\textcolor{blue}{-1})} &    \textcolor{purple}{+0} (\textcolor{orange}{-0})     &   -   &    \textcolor{purple}{+1} (\textcolor{orange}{-0})      & -      & -        \\    

    T1218.011-APT38                       &     7   & \multicolumn{1}{c|}{\textcolor{red}{+1} (\textcolor{blue}{-0})} &     6   & \multicolumn{1}{c|}{\textcolor{red}{+2} (\textcolor{blue}{-1})} &    \textcolor{purple}{+0} (\textcolor{orange}{-0})     &   \textcolor{purple}{+0} (\textcolor{orange}{-0})   &     -     &    -   &   \textcolor{purple}{+0} (\textcolor{orange}{-0})      \\    

    T1036.005-Aoqin Dragon                       &  5      & \multicolumn{1}{c|}{\textcolor{red}{+0} (\textcolor{blue}{-0})} &   5     & \multicolumn{1}{c|}{\textcolor{red}{+0} (\textcolor{blue}{-0})} &    \textcolor{purple}{+0} (\textcolor{orange}{-0})      &   \textcolor{purple}{+0} (\textcolor{orange}{-0})    &      -    &    \textcolor{purple}{+0} (\textcolor{orange}{-0})    &    -     \\                               
    T1071.001-APT32                       &  6      & \multicolumn{1}{c|}{\textcolor{red}{+1} (\textcolor{blue}{-0})} &   5     & \multicolumn{1}{c|}{\textcolor{red}{+1} (\textcolor{blue}{-0})} &    \textcolor{purple}{+0} (\textcolor{orange}{-0})      &   \textcolor{purple}{+0} (\textcolor{orange}{-0})    &      -    &    -    &    \textcolor{purple}{+1} (\textcolor{orange}{-0})     \\  

    T1047-APT32                      &  4      & \multicolumn{1}{c|}{\textcolor{red}{+0} (\textcolor{blue}{-0})} &   3     & \multicolumn{1}{c|}{\textcolor{red}{+0} (\textcolor{blue}{-0})} &    \textcolor{purple}{+0} (\textcolor{orange}{-0})      &   \textcolor{purple}{+0} (\textcolor{orange}{-0})    &      -    &    -    &    \textcolor{purple}{+1} (\textcolor{orange}{-0})     \\  
    T1204.002-EXOTIC LILY                       &  6      & \multicolumn{1}{c|}{\textcolor{red}{+1} (\textcolor{blue}{-0})} &   5     & \multicolumn{1}{c|}{\textcolor{red}{+1} (\textcolor{blue}{-0})} &    \textcolor{purple}{+0} (\textcolor{orange}{-0})      &   \textcolor{purple}{+1} (\textcolor{orange}{-0})    &      -    &    \textcolor{purple}{+0} (\textcolor{orange}{-1})    &    -     \\  

\midrule
Overall Precision                  &   1.000     & \multicolumn{1}{c|}{0.864} &    1.000    & \multicolumn{1}{c|}{0.811} &     1.000    &  0.929    &   0.981       &  1.000     &    0.963     \\
Overall Recall                  & 1.000 &   \multicolumn{1}{c|}{0.981}     & 1.000       & \multicolumn{1}{c|}{0.915}          & 0.981     &   1.000       &  1.000     &   0.981 & 1.000      \\
Overall F-1 Score                  &  1.000      & \multicolumn{1}{c|}{0.919}  & 1.000       & \multicolumn{1}{c|}{0.860} & 0.990        &   0.963   &    0.990      &    0.990   &0.981         \\
\bottomrule
\end{tabular}
\label{tab:RQ1CTIResult}  
\end{table*}

An attack technique generally consists of several attack behaviors, which are represented in an attack graph as a series of subgraphs consisting of interconnected system entities.

To accurately assess the performance of \SysName in extracting attack graphs from a single source, we compare the dynamic log-generated graphs and GPT-4 parsed CTI report-generated graphs with their respective manually constructed ground truth graphs to validate the accuracy of the attack knowledge graphs generated by our method.

For the attack graphs based on logs and static code sections, we used ten attack techniques labeled by hand, and by combining dynamic and static analyses, we labeled as many attack-related entities as possible and correlated the entities.
The results show in table \ref{logsingle} that \SysName achieves 93.8\% accuracy in recognizing entities and 91.4\% accuracy in recognizing relationships between entities.

Next, we manually analyzed the different attack techniques involved in eleven attack reports and labeled their related entities, inter-entity associations, and entity types. We compare the results with those obtained after analyzing them using GPT-4 and our designed prompt. As shown in Table \ref{singleCIT}, we recognize entities with 86.4\% accuracy, inter-entity associations with 81.1\% accuracy, and the minimum recognition rate is higher than 92.2\% for entity types.

The above experiments show that we can construct the technology graph effectively on a single resource.

\subsubsection{Effectiveness of the Subtechnique Graph Aggregation Module}\label{singlemerge}
For the same attack technique, information from different sources provides different perspectives on the attack knowledge.
Even for the same attack technique under the same data source, there are different code implementations and modes of operation, as well as descriptions from different CIT reports.
For example, the atomic red team provides 17 implementations for \texttt{T1547.001}, while the MITRE ATT\&CK species provides Procedure Examples for the \texttt{T1087} description from five reports.

We need to aggregate graphs of different techniques from the same source for the same attack technique in order to represent the attack more efficiently and concisely. We describe the specific algorithmic process for merging technique graphs from the same source in \S\ref{S51merghsingle}.

\begin{figure*}[hbtp!] 
\label{singlemerghegraph}
	\centering
	\subfigure[Entity Compression Rate Distribution]{
  \includegraphics[width=0.4\textwidth]{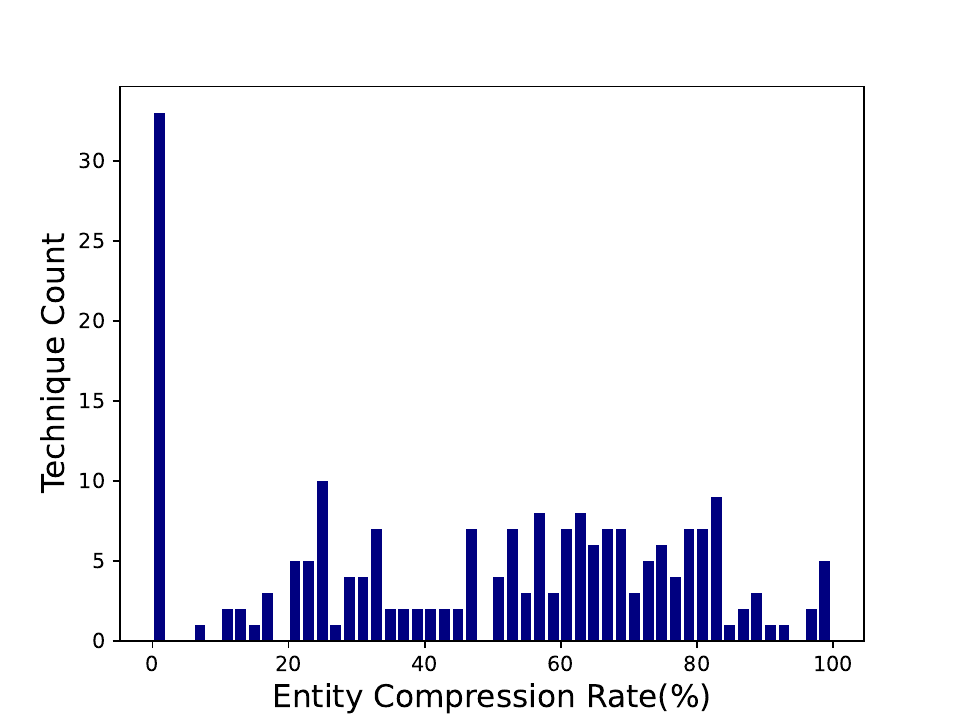}
		\label{Fig:audioEntity}
	}	
	\subfigure[Edge Compression Rate Distribution]{
  \includegraphics[width=0.4\textwidth]{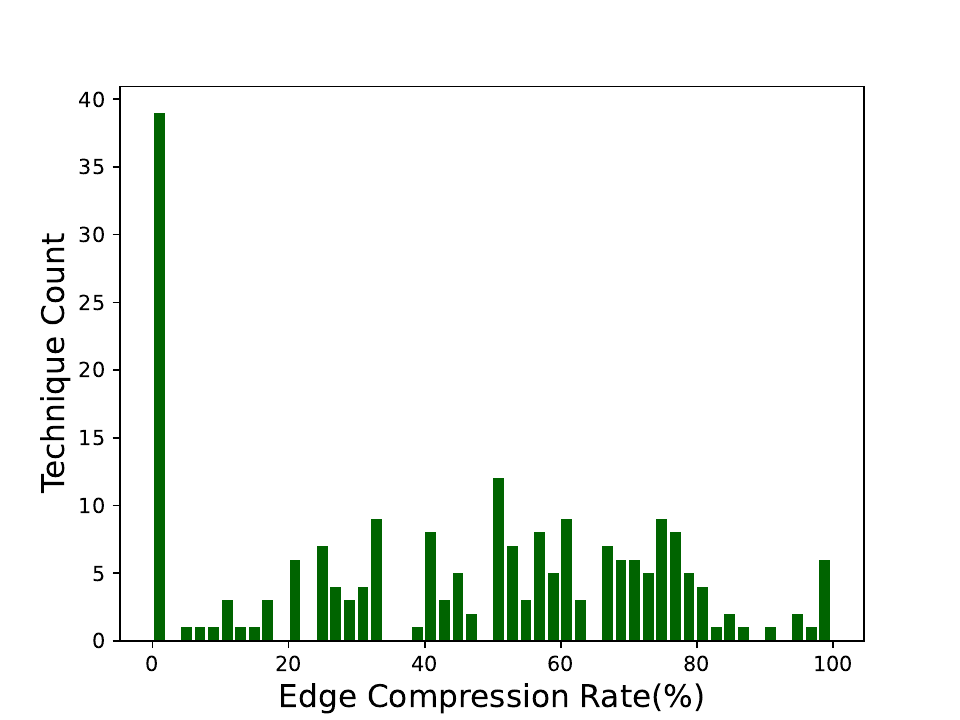}
		\label{Fig:audioedge}
	}	
	\caption{Compression ratio distribution of nodes and edges before and after aggregation for multiple procedure audit log knowledge graphs under each technology.}
	\label{Fig:audioCompressionRate}
\end{figure*}

\begin{figure*}[hbtp!] 
\setlength{\abovecaptionskip}{0pt}
\setlength{\belowcaptionskip}{0pt}
    \vspace{-0.1in}
	\centering	
  	\subfigure[Entity Compression Rate Distribution]{
  \vspace{-0.1in}
  \includegraphics[width=0.4\textwidth]{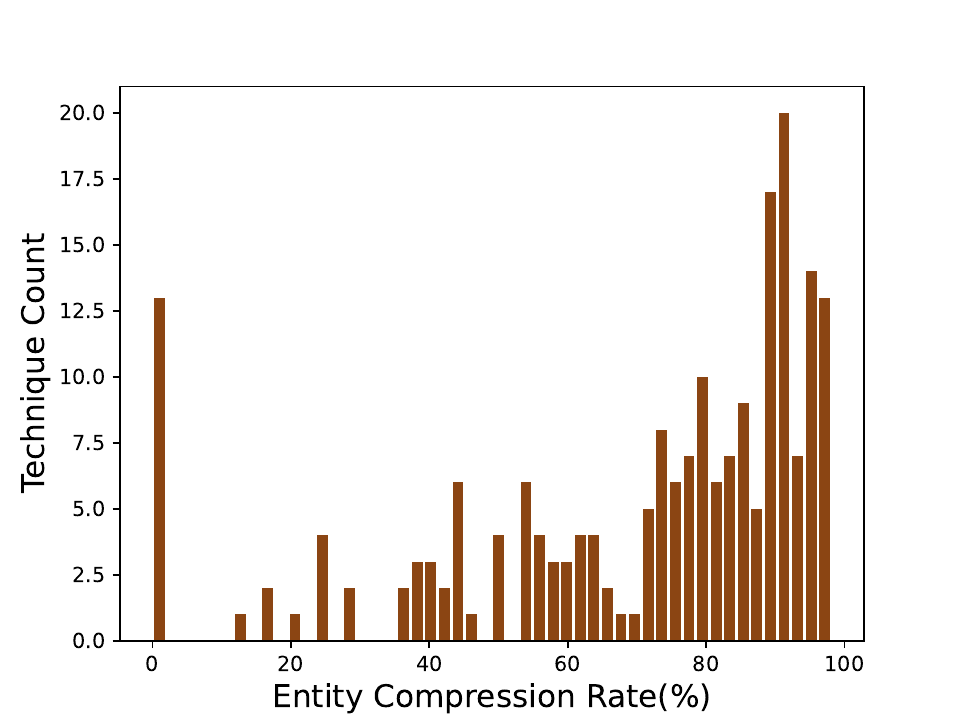}
		\label{Fig:CTIentity}
	}	
 	\subfigure[Edge Compression Rate Distribution]{
  \includegraphics[width=0.4\textwidth]{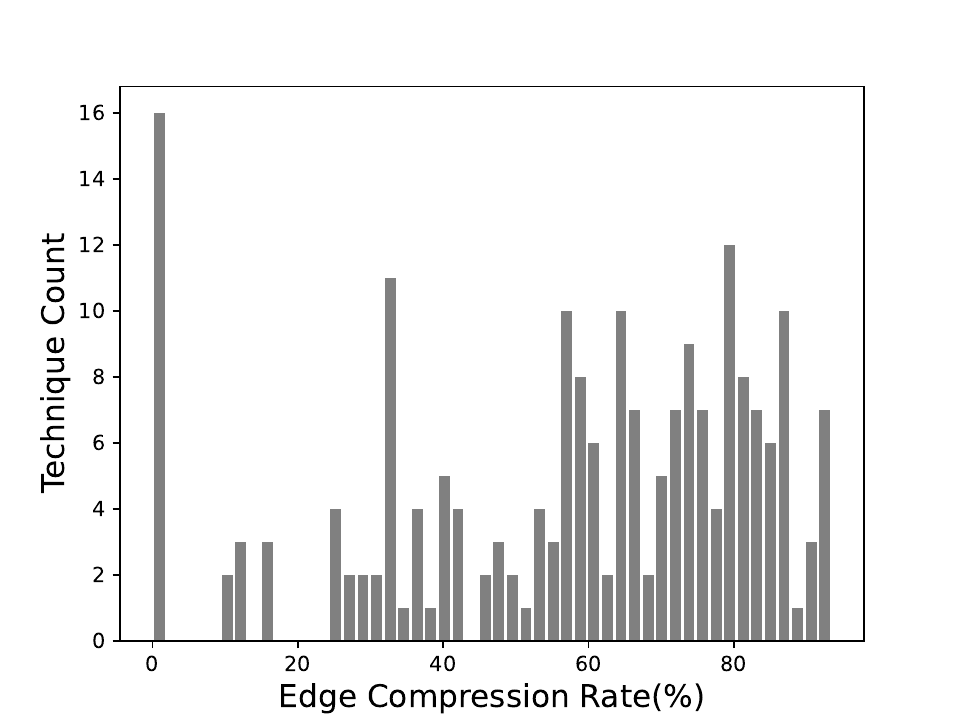}
		\label{Fig:CTIedge}
	}	
	\caption{Compression ratio distribution of nodes and edges before and after aggregation for multiple knowledge graphs on the same technology from different CIT reports.}
	\label{Fig: CTICompressionRate}
\end{figure*}

We have selected some of the results in Table~\ref{singlemergetableSome} for presentation, and the statistics of all the attack technique graphs are shown in Figure~\ref{Fig:audioCompressionRate} and Figure~\ref{Fig: CTICompressionRate}.
For log-based and code-sourced technical graphs, the average compression ratios of nodes and edges are respectively
47.41\% and 44.45\%. 
For the technical graph based on CTI report sources, the average compression ratio of nodes and edges is 69.97\% and 57.38\% respectively.
The compression rate of the audit log-based approach is lower than that of the report-based approach because the atomic red team provides only a single code implementation for some of the technologies, which cannot be further compressed. This is also the reason why we introduced threat reports, where a large number of threat reports can provide more implementation details about the same technology.

The effectiveness of our graph aggregation algorithm is demonstrated by comparing the number of procedures included in each attack technique, and the number of its nodes and edges, before and after aggregation.

\begin{table*}[h!]

\caption{The number of instances of attack techniques from different sources, and the retention ratio of entities and edges of their technique graphs before and after aggregation.
(We randomly select attack techniques collected from the atomic red team and CTI reports and count the audit log graph and report description graph information for each technique, as well as its post-aggregation data, to exemplify the effectiveness of our algorithm.)}
\label{singlemergetableSome}
\resizebox{\textwidth}{!}{%
\begin{tabular}{@{}l|ccccccc|ccccccc@{}}
\toprule
\multirow{2}{*}{Technique} & \multicolumn{3}{c}{Audit Log}                  & \multicolumn{4}{c|}{\SysName-Log}                                                                                                                       & \multicolumn{3}{c}{CTI Report}                 & \multicolumn{4}{c}{\SysName-CTI}                                                                                                                        \\ \cmidrule(l){2-15} 
                           & Procedure & Entity & \multicolumn{1}{c|}{Edge} & Entity & Edge & \begin{tabular}[c]{@{}c@{}}Entity \\ Retention(\%)\end{tabular} & \begin{tabular}[c]{@{}c@{}}Edge \\ Retention(\%)\end{tabular} & Procedure & Entity & \multicolumn{1}{c|}{Edge} & Entity & Edge & \begin{tabular}[c]{@{}c@{}}Entity \\ Retention(\%)\end{tabular} & \begin{tabular}[c]{@{}c@{}}Edge \\ Retention(\%)\end{tabular} \\ \midrule
T1003.001                  & 11        & 59     & \multicolumn{1}{c|}{49}   & 7      & 7    & 11.864                                                          & 14.286                                                        & 57        & 135    & \multicolumn{1}{c|}{143}  & 13     & 25   & 9.630                                                           & 17.483                                                        \\
T1018                      & 13        & 195    & \multicolumn{1}{c|}{421}  & 24     & 24   & 12.308                                                          & 5.701                                                         & 72        & 202    & \multicolumn{1}{c|}{188}  & 16     & 33   & 7.921                                                           & 17.553                                                        \\
T1021.002                  & 4         & 39     & \multicolumn{1}{c|}{37}   & 12     & 15   & 30.769                                                          & 40.541                                                        & 47        & 123    & \multicolumn{1}{c|}{118}  & 10     & 30   & 8.130                                                           & 25.424                                                        \\
T1040                      & 4         & 41     & \multicolumn{1}{c|}{39}   & 12     & 12   & 29.268                                                          & 30.769                                                        & 15        & 62     & \multicolumn{1}{c|}{61}   & 13     & 26   & 20.968                                                          & 42.623                                                        \\
T1036.003                  & 8         & 88     & \multicolumn{1}{c|}{95}   & 29     & 39   & 32.955                                                          & 41.053                                                        & 5         & 14     & \multicolumn{1}{c|}{9}    & 4      & 3    & 28.571                                                          & 33.333                                                        \\
T1059.001                  & 19        & 73     & \multicolumn{1}{c|}{56}   & 11     & 14   & 15.068                                                          & 25.000                                                        & 148       & 295    & \multicolumn{1}{c|}{211}  & 14     & 35   & 4.746                                                           & 16.588                                                        \\
T1548.002                  & 20        & 155    & \multicolumn{1}{c|}{143}  & 43     & 56   & 27.742                                                          & 39.161                                                        & 50        & 122    & \multicolumn{1}{c|}{94}   & 10     & 20   & 8.197                                                           & 21.277                                                        \\
T1546.002                  & 1         & 15     & \multicolumn{1}{c|}{18}   & 8      & 7    & 53.333                                                          & 38.889                                                        & 1         & 3      & \multicolumn{1}{c|}{2}    & 3      & 2    & 100.000                                                         & 100.000                                                       \\
T1090.003                  & 2         & 44     & \multicolumn{1}{c|}{51}   & 17     & 22   & 38.636                                                          & 43.137                                                        & 20        & 59     & \multicolumn{1}{c|}{52}   & 10     & 20   & 83.051                                                          & 61.538                                                        \\
T1615                      & 5         & 28     & \multicolumn{1}{c|}{27}   & 12     & 16   & 42.857                                                          & 59.259                                                        & 20        & 59     & \multicolumn{1}{c|}{52}   & 10     & 20   & 16.949                                                          & 38.462                                                        \\ \midrule
Average                    & 8.7       & 73.7   & \multicolumn{1}{c|}{93.6} & 17.5   & 21.2 & 23.745                                                           & 22.650                                                         & 43.5      & 107.4  & \multicolumn{1}{c|}{93}   & 10.3   & 21.4 & 28.816                                                          & 37.428                                                        \\ \bottomrule
\end{tabular}
}
\end{table*}

\subsubsection{Effectiveness of multi-source technical knowledge graphs}
To validate the effectiveness of cross-resource data integration, we compare the resulting final attack knowledge graph with the open-source CTI report parser AttacKG \cite{li2022attackg}, which is the state-of-the-art work on cross-report, graph-structure-based, attack technique awareness.

It is worth noting that some techniques could not be matched due to the non-corresponding technique granularity between the two parties, e.g., MITRE ATT\&CK defines \texttt{T1558.001} to \texttt{T1558.004}, but AttacKG analyzes it simply as \texttt{T1558}, which leads to the impossibility of comparison.
In addition, since we added two new data sources, static code, and dynamic logs, it resulted in a large gap between node information and types. 

Therefore, we found 65 matching techniques from the 179 techniques in AttacKG's work and compared their edges and nodes. Table \ref{comparewihtattackg} summarizes the statistical results of entity and relation extraction by \SysName and AttacKG in 6 techniques. The experimental results are in the statistics: Manual identifies manually labeled real data. AttacKG identifies the result data of AttacKG. \SysName identifies the result data by cross-source attack data. 

\begin{table*}[h!]
\centering  
\caption{The accuracy of \SysName and AttacKG for node and edge extraction compared to attack graphs after manually performing cross-source technique merging.(Nodes and edges for manual extraction of attack techniques, and \textcolor{red}{+false
\_positive} (\textcolor{blue}{-false\_negative}) for \SysName-generated graphs are presented separately, and finally the overall accuracy results.)}
\label{comparewihtattackg}
\begin{tabular}{@{}c|cccccc@{}}
\toprule
\multirow{2}{*}{Technique}             & \multicolumn{3}{c}{Node}                                                   & \multicolumn{3}{c}{Edge}                                           \\ \cmidrule(l){2-7} 
                                       & Manual               & \SysName                                             & \multicolumn{1}{c|}{AttacKG\cite{li2022attackg}}                                    & Manual               & \SysName                                             & AttacKG\cite{li2022attackg}                                              \\ \midrule
T1007                                  & 18                   & \textcolor{red}{+0} (\textcolor{blue}{-0})       & \multicolumn{1}{c|}{\textcolor{red}{+3} (\textcolor{blue}{-13})} & 18                   & \textcolor{red}{+0} (\textcolor{blue}{-0})       & \textcolor{red}{+3} (\textcolor{blue}{-11})           \\
T1027                                  & 9                    & \textcolor{red}{+1} (\textcolor{blue}{-0})       & \multicolumn{1}{c|}{\textcolor{red}{+4} (\textcolor{blue}{-4})}  & 9                    & \textcolor{red}{+1} (\textcolor{blue}{-0})       & \textcolor{red}{+10} (\textcolor{blue}{-3})           \\
T1070                                  & 8                    & \textcolor{red}{+1} (\textcolor{blue}{-0})       & \multicolumn{1}{c|}{\textcolor{red}{+9} (\textcolor{blue}{-2})}  & 7                    & \textcolor{red}{+1} (\textcolor{blue}{-0})       & \textcolor{red}{+30} (\textcolor{blue}{-2})           \\
T1120                                  & 7                    & \textcolor{red}{+2} (\textcolor{blue}{-0})       & \multicolumn{1}{c|}{\textcolor{red}{+0} (\textcolor{blue}{-4})}  & 7                    & \textcolor{red}{+2} (\textcolor{blue}{-0})       & \textcolor{red}{+1} (\textcolor{blue}{-5})            \\
T1137                                  & 7                    & \textcolor{red}{+0} (\textcolor{blue}{-0})       & \multicolumn{1}{c|}{\textcolor{red}{+0} (\textcolor{blue}{-4})}  & 7                    & \textcolor{red}{+0} (\textcolor{blue}{-0})       & \textcolor{red}{+1} (\textcolor{blue}{-6})            \\
T1547                                  & 10                   & \textcolor{red}{+1} (\textcolor{blue}{-0})       & \multicolumn{1}{c|}{\textcolor{red}{+5} (\textcolor{blue}{-2})}  & 9                    & \textcolor{red}{+1} (\textcolor{blue}{-0})       & \textcolor{red}{+25} (\textcolor{blue}{-1})           \\ 
\midrule
\multicolumn{1}{l|}{Overall Precision}  & 1.000                & 0.922                                            & \multicolumn{1}{c|}{0.588}                                       & 1.000                & 0.919                                            & 0.293                                                \\
\multicolumn{1}{l|}{Overall Recall}    & 1.000                & 1.000                                            & \multicolumn{1}{c|}{0.508}                                      & 1.000                & 1.000                                            & 0.509                                                \\
\multicolumn{1}{l|}{Overall F-1 Score} & 1.000                & 0.959                                            & \multicolumn{1}{c|}{0.545}                                       & 1.000                & 0.958                                            & 0.372                                                \\ \bottomrule
\end{tabular}
\end{table*}

It can be seen that the overall performance of \SysName is better than that of AttacKG. mainly because AttacKG simply extracts the rough textual information from the reports. 
Although AttacKG also uses a graph structure, the nodes are not recognized accurately enough, and many unarticulated nodes and edges appear. 
Due to the lack of information sources, the nodes related to the real attacks are not complete enough, but more inclined to produce a large number of false alarm nodes and edges. 
In addition, AttacKG produces many discrete subgraphs, isolated nodes, and self-looping edges.

As a result, \SysName's F-1 score is approximately twice as high as AttacKG's, both for node and edge comparison results. The experimental data suggests that for the construction of attack techniques, \SysName outperforms previous methods.

Similarly, the complete comparison results show that AttacKG has 43.76\% of the average number of nodes and 76.46\% of the number of edges of \SysName. Some of the results are shown in Table \ref{tab:attackg}.
By integrating the information across resources, we have greatly enriched the representation of individual attack techniques in terms of node information and their relationships.

In addition, we manually completed the merging of cross-source knowledge graphs and used this as ground truth to compare the two methods.
The results are shown in Table~\ref{comparewihtattackg}, where \SysName outperforms AttacKG in terms of both node and edge accuracy (92.2\% and 91.9\%, respectively).

\begin{table}[h]
\caption{Results of \SysName vs. AttacKG. (only some comparison data are shown, the Total Average is the average of all experiments). (NODES |V| AND EDGES |E|, Ratio represents AttacKG's proportion of \SysName).}
\label{tab:attackg}
\resizebox{0.45\textwidth}{!}{%
\begin{tabularx}{0.5\textwidth}{l*{4}{>{\centering\arraybackslash}X}}
\toprule
\textbf{Technique} & \multicolumn{2}{c}{\textbf{\SysName}} & \multicolumn{2}{c}{\textbf{AttacKG\cite{li2022attackg}}} \\
& {$|V|$} & {$|E|$} & {$|V|$/Ratio} & {$|E|$/Ratio} \\ 
\midrule
T1007 & 18 & 18 & {8/44.44\%} & {10/55.56\%} \\
T1010 & 11 & 10 & {2/18.18\%} & {1/10.0\%}   \\
T1012 & 14 & 15 & {9/64.29\%} & {13/86.67\%} \\
T1016 & 25 & 25 & {8/32.0\%}  & {23/92.0\%}  \\
T1018 & 28 & 28 & {8/28.57\%} & {13/46.43\%} \\
T1046 & 11 & 11 & {4/36.36\%} & {4/36.36\%}  \\
T1047 & 13 & 15 & {16/123.08\%} & {20/133.33\%} \\
T1048 & 6  & 5  & {3/50.0\%} & {3/60.0\%}   \\
T1049 & 18 & 19 & {7/38.89\%} & {10/52.63\%} \\
T1055 & 12 & 12 & {9/75.0\%} & {17/141.67\%} \\
T1070 & 9  & 8  & {15/166.67\%} & {35/437.5\%} \\
T1057 & 12 & 12 & {10/83.33\%} & {23/191.67\%} \\
T1113 & 15 & 17 & {8/53.33\%} & {12/70.59\%} \\
T1115 & 11 & 14 & {2/18.18\%} & {1/7.14\%} \\
T1070 & 9  & 8 & {15/166.67\%} & {35/437.5\%}  \\ 
\bottomrule
Total Average & -  & - & {-/43.76\%} & {-/76.46\%}  \\ 
\bottomrule
\end{tabularx}
}
\end{table}

\subsubsection{Case Study.}
This subsection discusses how \SysName can be employed in real-world security tasks through a case study. Specifically, \SysName utilizes multi-sourced knowledge to refine the representation of an attack technique, thereby helping to detect and identify different variants of the executed attack technique and their constituent attack chains in real-world systems.
Furthermore, \SysName can supplement the understanding of attack techniques among security professionals, supporting tasks related to attack reconstruction and replication.

\textbf{\SysName for Attack Technique Reconstruction.} Our final cross-resource merged technology knowledge graph is a fine-grained reconstruction of specific attack techniques, which contains structured information from multiple sources and is richer and more complete.
Compared to previous work, cross-resource knowledge of attack techniques can bridge the knowledge gap between real attacks and CTI reports. We not only analyze the attack techniques across reports but also add static code analysis and dynamic log analysis, which can restore the details of the attack techniques at a finer granularity and more completely, which can be used to achieve the attack reconstruction goals.

\begin{figure*}[h!] 
    \centering
\includegraphics[width=0.75\textwidth]{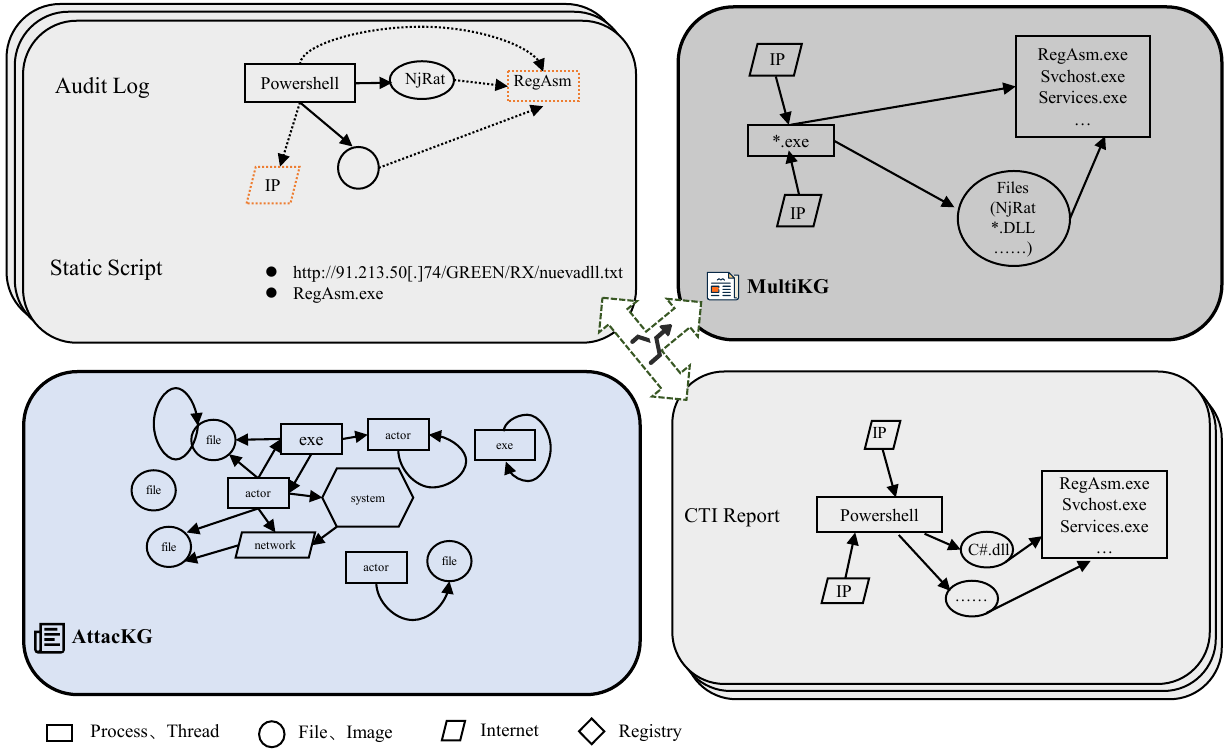}
        \caption{Cross-Source Knowledge Fusion Based on Multi-Source Threat Intelligence: A Case Study of \texttt{T1055.001}.}
    \label{fig:casestudyT1055.001}
\end{figure*}

Specifically, as shown in Figure~\ref{fig:casestudyT1055.001}, compared to the single-source-based approach, \SysName's ability to summarize attack knowledge across resources and achieve reconstruction and enhancement effects on attacks in audit logs, static code, and threat reports can help to reconstruct the involved attacks and detect them, and the cross-resource information can also enhance the detection of attack variants.

Based on the entities involved in our attack knowledge graph, inter-entity relationships, etc., we can infer the execution environment, execution path, and corresponding operations of the attack techniques, which in turn can realize their reconstruction work.

\textbf{\SysName for Attack Technique Detection.} Single-source and coarse-grained data make it difficult to reflect the details of a real attack execution due to the limited amount of information and limited coverage of different variants of attack techniques.
We can effectively improve the detection of attack techniques and their various related variants by collecting and integrating fine-grained attack technique data across resources and generalizing the information.
Thus, \SysName can enhance any detector that relies on the Tactics-Techniques-Procedures (TTP) heuristic to provide a richer and complete attack knowledge base for downstream tasks.

\begin{figure*}[h!] 
    \centering
\includegraphics[width=0.85\textwidth]{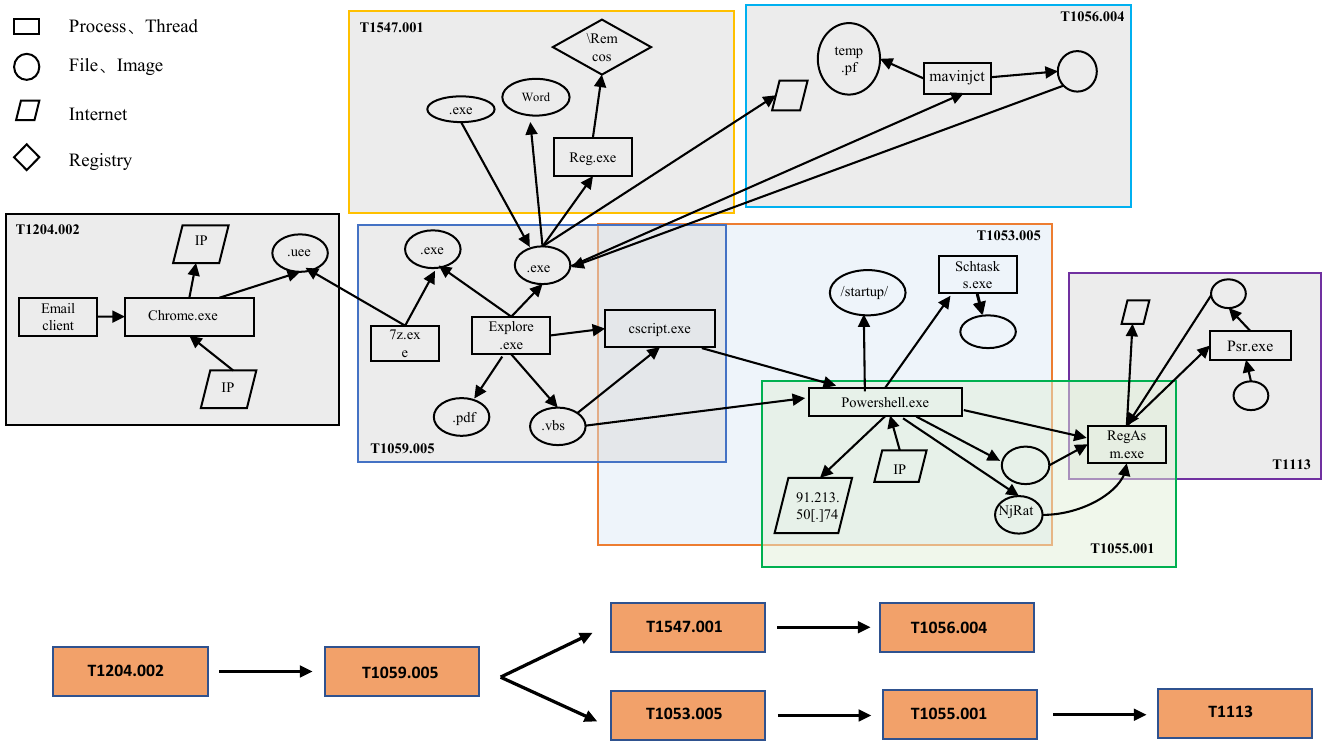}
        \caption{\SysName for Attack Detection (APT-C-36)}
    \label{fig:casestudyDetection}
\end{figure*}

As shown in Figure~\ref{fig:casestudyDetection}, we can use the attack knowledge graph after integrating cross-source information to identify individual attack techniques and their variants in real attack scenarios by methods such as graph matching and construct attack chains by combining attack techniques based on timestamps and public nodes.

%% file: 9_Related_Wrok.tex
\section{Related Wrok}

\subsection{Threat Intelligence Gathering}

Cyber threat intelligence (CTI) plays a vital role in security warfare to keep up with the rapidly evolving landscape of cyber attacks~\cite{mu2018understanding,gao2021system}.
To facilitate CTI knowledge exchange and management, the security community has standardized open formats (e.g., OpenIoC~\cite{OpenIOC}, STIX~\cite{STIX}, and CybOX~\cite{CybOX}) to describe Indicators of Compromises (IoCs).
Though structured and machine-readable, such intelligence lacks semantic information about how entities/IoCs interact to form the kill chain.

Attack techniques cannot be accurately represented if only by unconnected isolated IoCs (such as the names and types of malicious files, the number and names of malicious processes, etc.), and are easily evaded by attackers.
In addition, existing graph construction methods rely on a small number of manually labeled datasets, lack sufficient security knowledge to learn~\cite{zhao2018document,kodelja2019exploiting}, and require the expert design of natural language processing or graph matching models, which greatly limits the accuracy and validity of the extracted results and challenges security personnel's workload and technical background.
In this paper, we try to improve this technical gap by using LLM to aggregate technical-level knowledge from CTI reports and perform graphical construction.
LLM uses a large amount of open-knowledge data in the pre-training process, which does not require the preparation of sufficient datasets and well-designed network structures and can achieve good contextual understanding and knowledge reasoning capability only through few-shot learning or zero-shot learning, so we perform attack knowledge graph construction based on LLM~\cite{touvron2023llama, brown2020language}.

Poirot~\cite{Milajerdi2019} utilizes manually extracted CTI query graphs for intrusion detection in a provenance graph constructed based on system logs. 
However, it does not consider the huge information gap between CIT reports and real attack logs, which leads to poor results for matching and detection.
In \SysName, we introduce system audit logs to bridge the gap between the two, and we also introduce a more practical static code analysis to help refine the selection of events in dynamic log graphs, by which we construct a more comprehensive and close-to-real attack knowledge graph by using the above three attack knowledge sources.



\subsection{LLMs for Cyber Security}
Currently, In recent years, many studies have explored the practical applications of LLMs in the field of cyber security.
Maxime et al. \cite{wursch2023llms} LLMs to extract relevant knowledge entities from cybersecurity-related texts. AttacKG+~\cite{zhang2024attackgboostingattackknowledgegraph} leverages LLM to convert CTI reports into structured representations and then parses entities and relationships from which attack techniques are identified and summarized.
Maxime uses LLM to improve the effectiveness of existing extractors by extracting relevant knowledge entities from cybersecurity-related text. KGV~\cite{wu2024kgvintegratinglargelanguage} proposed a Cyber Threat Intelligence (CTI) quality assessment framework based on LLMs for fact-checking knowledge graphs constructed from paragraphs. \cite{wang2024sandsmansionsenablingautomatic} can autonomously build multi-stage cyberattack plans based on Cyber Threat Intelligence (CTI) reports, construct the emulation infrastructures, and execute the attack procedures.

\subsection{Attack Investigation}
\subsubsection{Anomaly-based attack investigation}
Threat alerting based on anomalies is a more common approach to network security.
EIGER utilizes the Indicator of Compromise (IOC) extracted from CTI reports for attack alerting, while NODOZE~\cite{hassan2019nodoze} constructs a database of event frequencies that is analyzed to derive anomaly scores and give an alert event dependency graph.

Detection based on a single IOC is prone to many false positives, and the number of alerts to be handled will be much larger than the number of analysts. Attack analysis based on normal event frequencies, on the other hand, is likely to be bypassed by the latest mimic attacks and variants of attack techniques.
\subsubsection{Attack Investigation Based on Cyber Threat Intelligence Knowledge}

Recent advancements of causality analysis in provenance graph~\cite{king2003backtracking,king2005enriching,pohly2012hi,bates2015trustworthy} have enabled security investigators to detect cyber attacks with multiple stages.
However, audit logs monitor general-purpose system activities and thus lack the knowledge of high-level behaviors~\cite{zengwatson}.
In most cases, analysts act as the backbone in SOCs (security operations centers) to correlate various attack stages through reviewing numerous system logs~\cite{vanede2020deepcase}.
Unfortunately, as the volume of audit data is typically overwhelming even after reducing noisy logs irrelevant to attacks~\cite{lee2013loggc,xu2016high,liu2018towards,tang2018nodemerge,michael2020forensic,hassan2020we}, it is infeasible to manually analyze cyber threats directly based on audit logs~\cite{gao2018aiql,gao2018saql}.

CTI has proved to be valuable information in navigating audit logs and pinpointing behaviors associated with known attacks~\cite{MomeniMilajerdi2019,Milajerdi2019,hossain2020combating,Hassan2020,gao}.
For example, RapSheet~\cite{Hassan2020} matches audit logs against a TTP knowledge base from MITRE and correlates TTPs in logs, and constructs a TPG (tactic provenance graph) to recover \textit{kill chains} in APT campaigns.
We envision that the deployment of \SysName can enhance any detector that relies on TTP heuristics by providing a fully automated solution for building attack knowledge graphs from multiple sources.

%% file: 10_Conclusion.tex
\section{Conclusion}
In this paper, we propose a feasible scheme to extract and process attack knowledge across information sources and do a fusion representation of it to build fine-grained attack knowledge graphs.
First, we utilize system audit logs and static code analysis to construct a technology-level dynamic attack graph. Then, we utilize LLM to parse the threat report knowledge provided by MITRE ATT\&CK to construct a technology-level attack graph. Finally, we aggregate same-source attack graphs and merge cross-source attack graphs to generate the final attack knowledge graph, which improves the accuracy and completeness of the attack knowledge graph representation.

We implemented and deployed the \SysName system and evaluated it with 1015 real attack techniques and corresponding 9006 attack technique intelligence from real-world CTI reports.
The results show that \SysName can accurately extract attack graphs from different data sources, aggregate them, and efficiently summarize and merge cross-source technique-level attack knowledge to produce an accurate and complete knowledge graph of attack techniques.


%% file: 11_Appendix.tex
\newpage 
\section{Cases of Attack Knowledge Graphs Constructed by \SysName}\label{APP_A}
\begin{figure}[h!]
    \centering
    \includegraphics[width=0.5\textwidth]{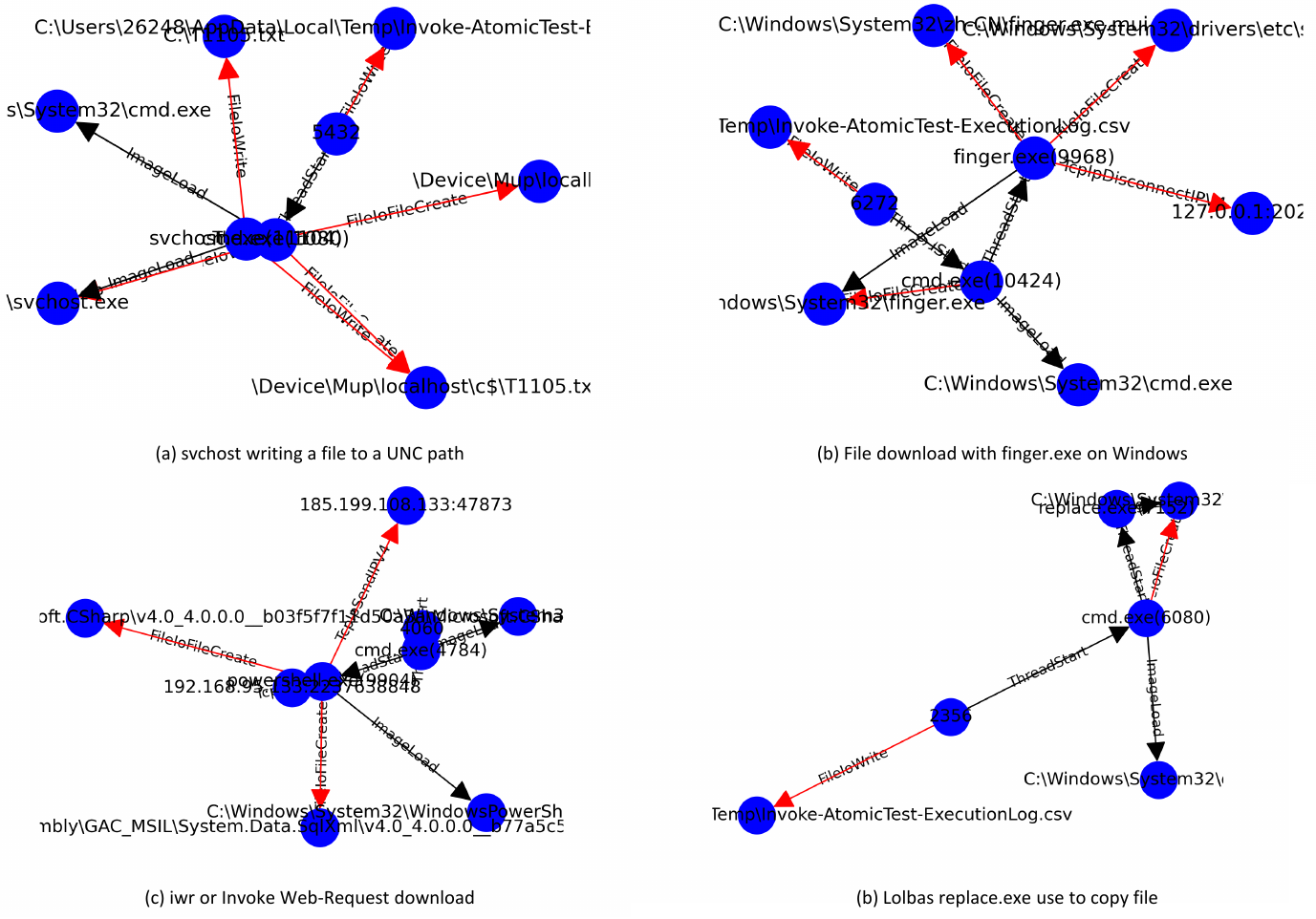}
    \caption{Example of Audit Log Based Attack Graph Visualization Constructed by \SysName (\texttt{T1105})}
    \label{fig:APPauditlog}
\end{figure}

\begin{figure}[h!]
    \centering
    \includegraphics[width=0.5\textwidth]{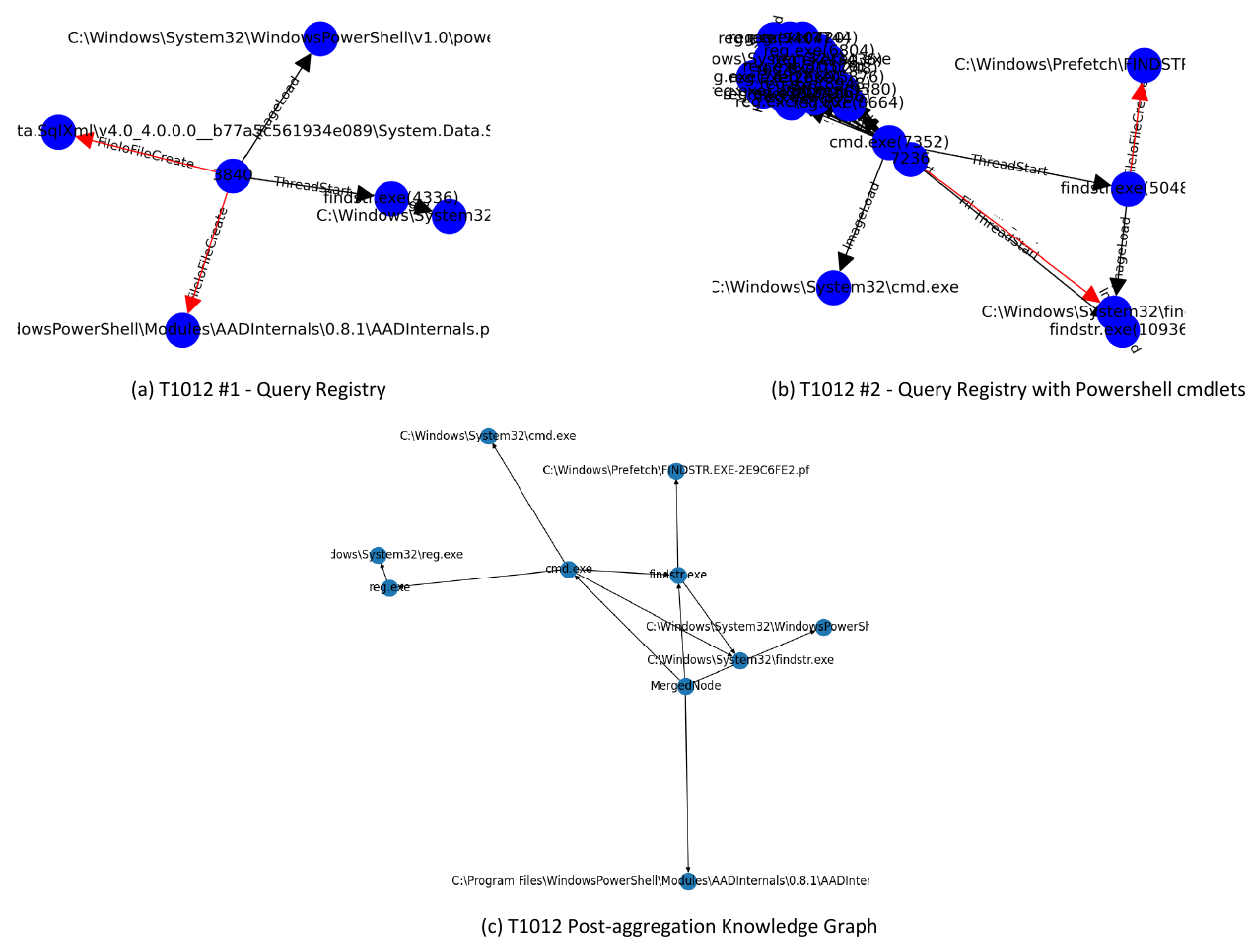}
    \caption{Example of Post-Aggregation Attack Knowledge Graph Based on Audit Log (\texttt{T1012})}
    \label{fig:APPauditlog2}
\end{figure}

\begin{figure}[h!]
    \centering
\includegraphics[width=0.5\textwidth]{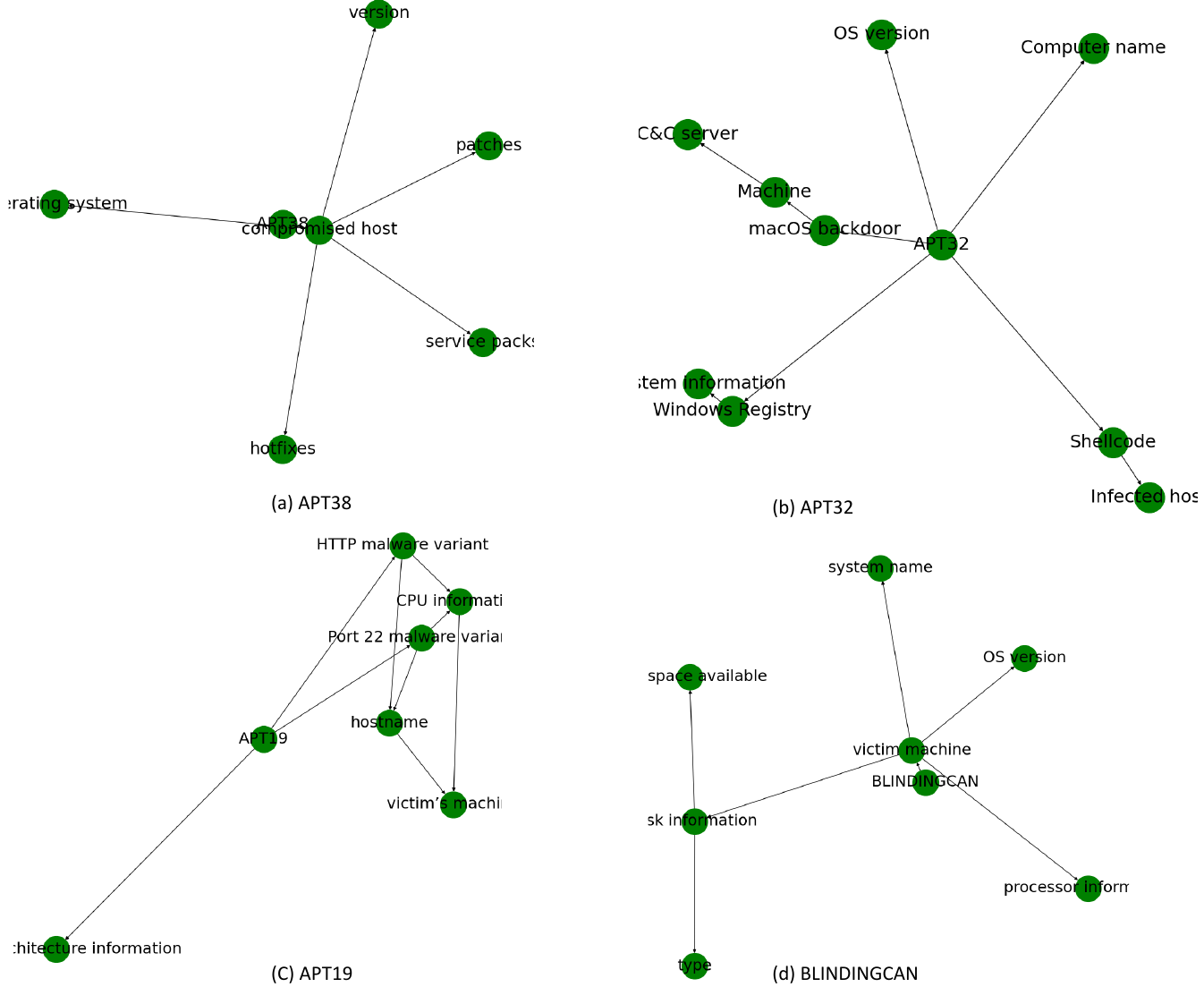}
    \caption{Attack Knowledge Graph Based on LLM Parsing (\texttt{T1082} technique from four different CTI reports, respectively)}
    \label{fig:APPcti1}
\end{figure}

\begin{figure}[h!]
    \centering
    \includegraphics[width=0.5\textwidth]{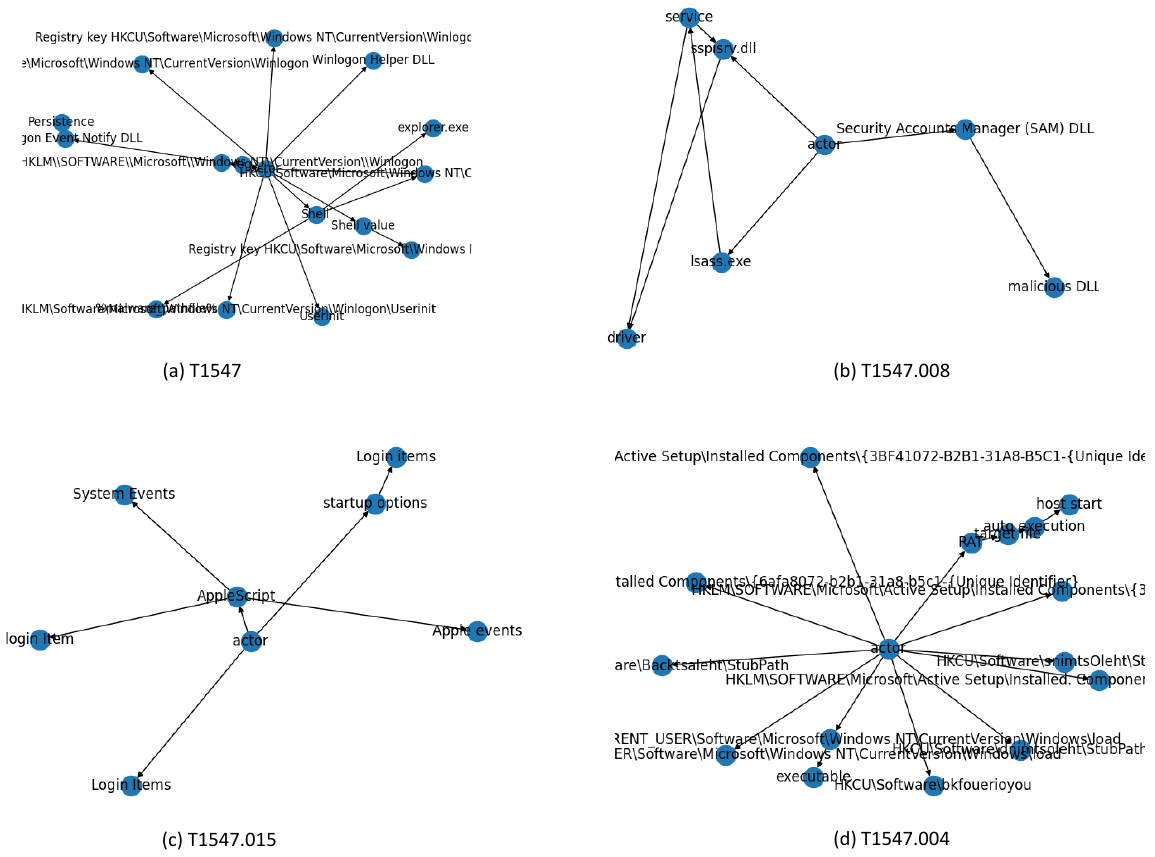}
    \caption{Example of Attack Knowledge graphs Based on LLM Parsing and Processed by the Aggregation Algorithm}
    \label{fig:APPcit2}
\end{figure}

\begin{figure}[h!]
    \centering
    \includegraphics[width=0.5\textwidth]{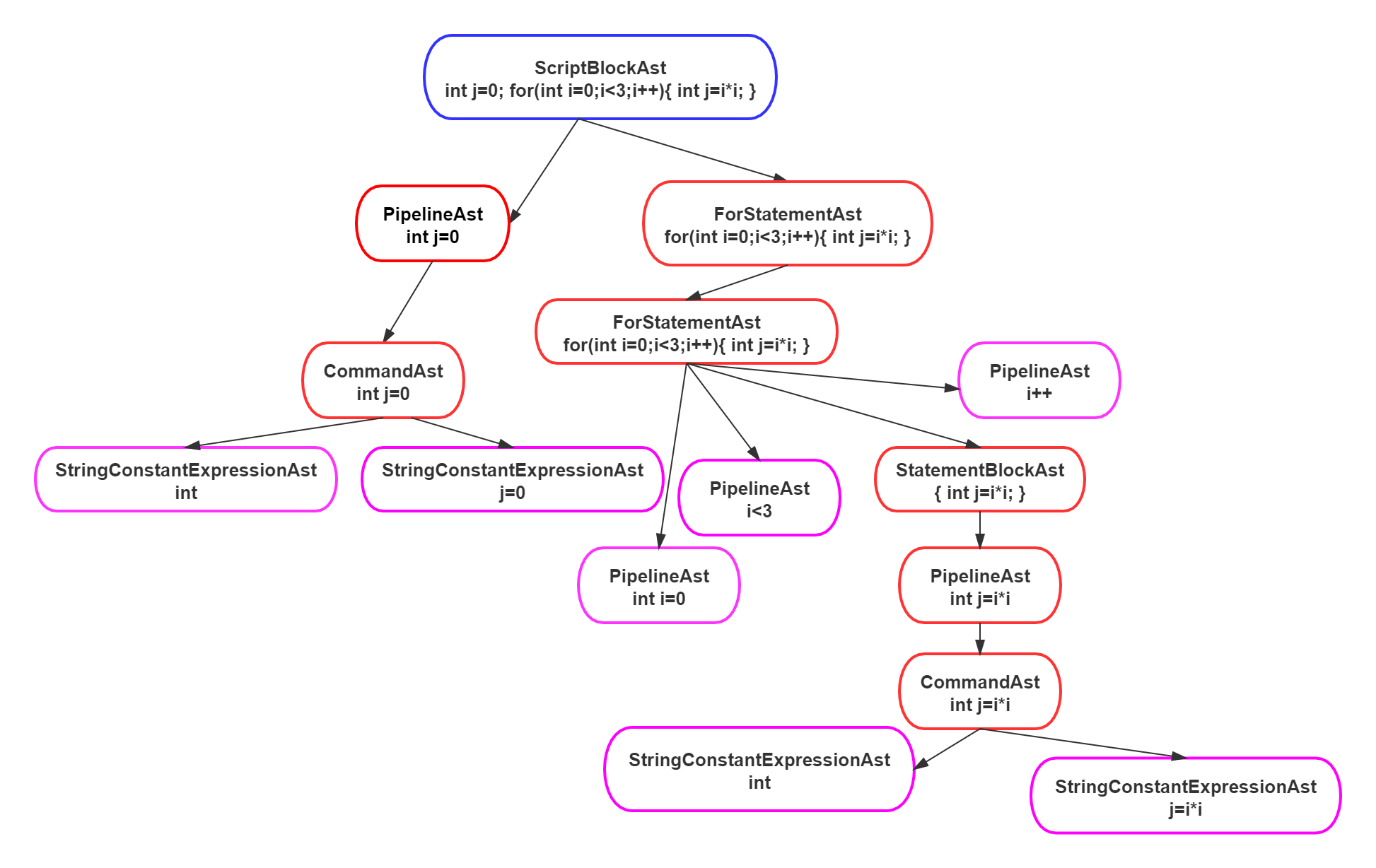}
    \caption{Example Graph of Abstract Syntax Tree Visualization}
    \label{fig:APPAST}
\end{figure}

\FloatBarrier 
Appendix \ref{APP_A} provides a series of examples demonstrating the visualization of attack knowledge graphs extracted from cyber-attack events described in attack knowledge from various sources, the example sources being system audit logs, CTI reports, and static code, respectively.
Figure~\ref{fig:APPauditlog} and Figure~\ref{fig:APPauditlog2} show the attack knowledge graphs extracted based on audit logs and their comparison before and after aggregation, respectively.
Figure~\ref{fig:APPcti1} and Figure~\ref{fig:APPcit2} show the attack knowledge graph of CIT reports extracted based on LLM, and its result after aggregation, respectively.
Figure~\ref{fig:APPAST} shows an example presentation of a simple AST.

